\documentclass[aps,prl,twocolumn,showpacs,superscriptaddress,longbibliography]{revtex4-2}
\usepackage{amsmath,amssymb,graphicx}
\usepackage{graphicx,epstopdf}
\usepackage{gensymb}
\usepackage[dvipsnames]{xcolor}
\usepackage[T1]{fontenc}
\usepackage[utf8]{inputenc}
\newcommand{\be}{\begin{equation}}
	\newcommand{\ee}{\end{equation}}
\newcommand{\bea}{\begin{eqnarray}}
	\newcommand{\eea}{\end{eqnarray}}
\newcommand{\bse}{\begin{subequations}}
	\newcommand{\ese}{\end{subequations}}

\usepackage{multibib}
\usepackage{color}
\usepackage[colorlinks,bookmarks=false,citecolor=darkblue,linkcolor=red,urlcolor=blue]{hyperref}

\definecolor{darkred}{rgb}{0.7,0.0,0.0}

\definecolor{darkblue}{rgb}{0,0.02,0.45}

\definecolor{darkgreen}{rgb}{0.02,0.45,0.0}

\definecolor{violet}{rgb}{0.8,0.2,0.6}

\begin{document}
\title{Absence of magnetic order and emergence of unconventional fluctuations in $J_{\rm eff} =1/2$ triangular lattice antiferromagnet YbBO$_3$}

\author{K. Somesh}
\author{S. S. Islam}
\author{S. Mohanty}
\affiliation{School of Physics, Indian Institute of Science Education and Research Thiruvananthapuram-695551, India}
\author{G. Simutis}
\email{gediminas.simutis@psi.ch}
\affiliation{Laboratory for Neutron and Muon Instrumentation, Paul Scherrer Institut, CH-5232 Villigen PSI, Switzerland}
\affiliation{Department of Physics, Chalmers University of Technology, SE-41296 G\"oteborg, Sweden}
\author{Z. Guguchia}
\author{Ch. Wang}
\affiliation{Laboratory for Muon Spin Spectroscopy, Paul Scherrer Institut, Villigen PSI, Switzerland}
\author{J. Sichelschmidt}
\author{M. Baenitz}
\affiliation{Max Planck Institute for Chemical Physics of Solids, Nöthnitzer Strasse 40, 01187 Dresden, Germany}
\author{R. Nath}
\email{rnath@iisertvm.ac.in}
\affiliation{School of Physics, Indian Institute of Science Education and Research Thiruvananthapuram-695551, India}
\date{\today}

\begin{abstract}
We present the ground state properties of a new quantum antiferromagnet YbBO$_3$ in which the isotropic Yb$^{3+}$ triangular layers are separated by a non-magnetic layer of partially occupied B and O(2) sites. The magnetization and heat capacity data establish a spin-orbit entangled effective spin $J_{\rm eff} = 1/2$ state of Yb$^{3+}$ ions at low temperatures, interacting antiferromagnetically with an intra-layer coupling $J/k_{\rm B} \simeq 0.53$~K. The absence of oscillations and a $1/3$ tail in the zero-field muon asymmetries rule out the onset of magnetic long-range-order as well as spin-freezing down to 20~mK.
An anomalous broad maximum in the temperature dependent heat capacity with a unusually reduced value and a broad anomaly in zero-field muon depolarization rate centered at $T^*\simeq 0.7 \frac{J}{k_{\rm B}}$ provide compelling evidence for a wide fluctuating regime ($0.182 \leq T/J \leq 1.63$) with slow relaxation. We infer that the fluctuating regime is a universal feature of a highly frustrated triangular lattice antiferromagnets while the absence of magnetic long-range-order is due to perfect two-dimensionality of the spin-lattice protected by non-magnetic site disorder.
\end{abstract}

\maketitle

Frustrated magnets are a special class of systems where the ground state degeneracy leads to captivating low temperature properties. One such emergent phenomenon is the quantum spin liquid (QSL), a disordered state characterized by fractionalized excitations, quantum entanglement, and absence of magnetic long-range-order (LRO)~\cite{Balents199,*Broholmeaay0668}.
The quest for this exotic phase has escalated remarkably since Anderson's proposal of resonating valence bond model, a prototype of QSL in spin-$1/2$ triangular lattice antiferromagnets (TLAFs)~\cite{Anderson153}. However, the subsequent theoretical studies revealed three-sublattice N\'eel order as the real ground state for Heisenberg TLAFs~\cite{Singh1766,*Capriotti3899}.
On a pragmatic point of view, the real materials are often plagued with certain degree of perturbations such as exchange anisotropy, couplings beyond nearest-neighbor (NN), structural disorder etc that may alter the actual ground state and pave the way even for richer physics~\cite{Watanabe034714,Kimchi031028,Li014426,Maksimov021017}. A prominent manifestation of such effects is the broad fluctuating regime with slow dynamics observed in TLAFs (Na,K,H)CrO$_2$ in low temperatures~\cite{Somesh104422,Olariu167203}. Indeed, spin-$1/2$ TLAFs with above perturbations are recently perceived as an ideal host for gapped/gapless QSL and other exotic pahses~\cite{Kaneko093707,Li224004}.

Rare-earth ($4f$) based TLAFs, mainly with ytterbium (Yb$^{3+}$) set a new platform to explore non-trivial phases of matter. Here, the interplay of strong spin-orbit coupling (SOC) and non-cubic crystal electric field (CEF) lead to Kramers doublet with an effective $J_{\rm eff} = 1/2$ ground state.
Moreover, the strong SOC is expected to induce anisotropic exchange interactions, which is proposed to be an ingredient in stabilizing QSL state~\cite{Li035107}. Further, disorder in the frustrated magnets is believed to destroy the states that are stabilized in disorder-free compounds and concedes various fascinating ground states~\cite{Furukawa077001}. For instance, recently, it was found that structural disorder can trigger bond randomness and promote QSL like states~\cite{Kimchi031028,Watanabe034714}.
The most celebrated compounds in this category are Yb(Mg,Zn)GaO$_4$ where disorder due to site mixing of Ga$^{3+}$ and Mg$^{2+}$/Zn$^{2+}$ leads to randomized interactions which apparently forbids LRO and results highly fluctuating moments~\cite{Li167203,Li16419,Li107202,Zhu157201,Zhang031001,Ma087201,Liu041040}. This opens up a new paradigm for realizing randomness induced non-trivial states that mimic QSL~\cite{Ma224433,Paddison117,Li097201,Bchmidt214445}. The disorder free chalcogenides (Na,Cs,Li)Yb$X_2$ ($X=$~O, S, and Se) featuring triangular lattice also show signature of QSL in zero-field and field induced transition in higher fields~\cite{Baenitz220409,Ranjith180401,Bordelon1058,*Ding144432,Ranjith224417,Guo064410,Bordelon224427,Zhang085115,*Dai021044,Sarkar241116}. Though, no LRO is detected down to 50~mK in another disorder-free TLAF Ba$_3$Yb(BO$_3$)$_3$ but the dominant interaction is found to be long-range dipole-dipole coupling rather than exchange interactions~\cite{Zeng045149,BagL220403}.

\begin{figure}
	\includegraphics[width=3.1 in, height= 2.4 in]{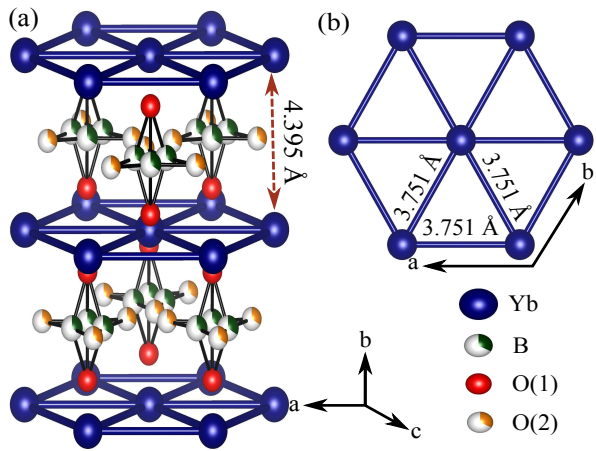}
	\caption{\label{Fig1} (a) Crystal structure of YbBO$_3$ showing triangular Yb$^{3+}$ sheets separated by a layer of B and O atoms. The O(1) site is fully occupied while O(2) and B sites are only occupied 33\% each. (b) A section of the Yb$^{3+}$ layer with equilateral triangles.}
\end{figure}

Herein, we report a comprehensive study of a new TLAF YbBO$_3$ (hexagonal, $P6_{3}/m$) with partial non-magnetic site-disorder. The magnetic Yb$^{3+}$ ions fully occupy the regular triangular sites and are free from any sort of magnetic site-disorder (see Fig.~\ref{Fig1})~\cite{Chadeyron261}. One of the oxygen site [O(2)] out of two and the only one boron site are 33\% occupied each. The triangular layers are separated by a layer of these partially occupied B and O(2) sites. We choose YbBO$_3$ in order to investigate how the non-magnetic site disorder affects the local environment of Yb$^{3+}$ and hence the ground state.
We illustrate that the Yb$^{3+}$ ions form Kramers doublet with effective spin $J_{\rm eff} = 1/2$ at low temperatures. Despite a relatively large Curie-Weiss temperature $\theta_{\rm CW} \simeq -0.8$~K, magnetic LRO is forbidden down to 20~mK. Intriguingly, the $\mu$SR measurements reveal a broad low temperature fluctuating regime with slow dynamics. These features are attributed to the combined effects of magnetic frustration and non-magnetic site disorder.
Polycrystalline sample of YbBO$_{3}$ was synthesized via the conventional solid-state reaction technique. The phase purity and crystal structure [hexagonal, $P6_{3}/m$ (No. 176)] of the compound were confirmed from powder x-ray diffraction [see the Supplementary Material (SM) for details~\cite{supplementary}]~\cite{Chadeyron261}. Magnetization ($M$) was measured using a SQUID magnetometer (MPMS-3, Quantum Design) down to 0.4~K using a $^3$He (iHelium3, Quantum Design Japan) attachment to the MPMS. Heat capacity [$C_{\rm p}(T)$] measurement was performed on a piece of pellet using a PPMS (Quantum Design). For measurements down to 0.4~K, a $^3$He attachment was used in PPMS. The $\mu$SR measurement was performed at the S$\mu$S muon source at Paul Scherrer Institute using a combination of two spectrometers (GPS and HAL) down to 20~mK in zero-field. The details of the $\mu$SR experiment is described in the SM\cite{supplementary}.
\begin{figure}
	\includegraphics[width=\columnwidth]{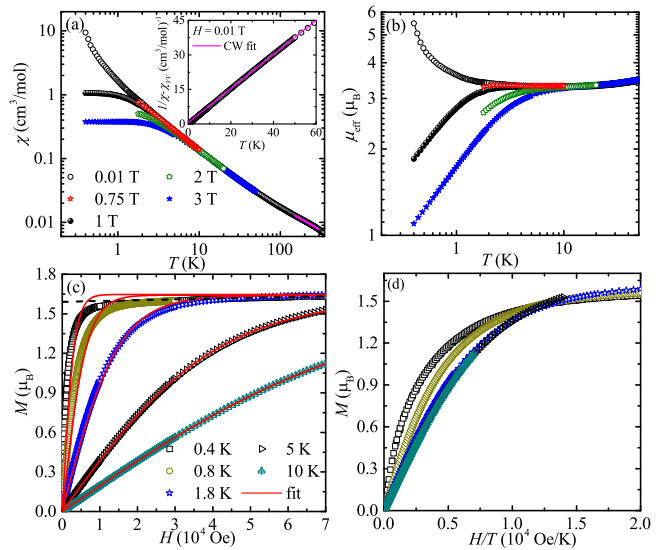}
	\caption{\label{Fig2} (a) $\chi(T)$ of YbBO$_3$ measured in various applied magnetic fields. Inset: The low-temperature $\chi(T)$ data (after subtracting $\chi_{\rm VV}$) along with the CW fit for $H = 0.01$~T. (b) Effective moment ($\mu_{\rm eff}$) as a function of $T$. (c) $M$ vs $H$ measured at different temperatures along with the Brillouin fits (solid lines). The horizontal dashed line marks the Van-Vleck contribution. (d) $M$ vs $H/T$ to visualize the scaling of magnetization curves in the correlated regime.}
\end{figure}
Magnetic susceptibility $\chi~[\equiv M/H]$ as a function of temperature in different applied fields ($H$) is depicted in Fig.~\ref{Fig2}(a). $\chi(T)$ increases monotonically as the temperature is lowered without showing any signature of magnetic long-range-order (LRO) down to 0.4~K, a possible finger print of disorder ground state.
Above 150~K, the inverse susceptibility ($1/\chi$) for $H = 0.01$~T could be fitted well by $\chi(T)=\chi_0+\frac{C}{T-\theta_{\rm CW}}$, where $\chi_0$ is the $T$-independent susceptibility and the second term is the Curie-Weiss (CW) law. The fit yields $\chi_0 \simeq 6.3 \times 10^{-4}$~cm$^3$/mol, a high-$T$ effective moment $\mu_{\rm eff}^{\rm HT}$ $[=\sqrt{(3k_{\rm B}C/N_{\rm A}\mu_{\rm B}^{2})}$, where $C$, $k_{\rm B}$, $N_{\rm A}$, and $\mu_{\rm B}$ are the Curie constant, Boltzmann constant, Avogadro’s number, and Bohr magneton, respectively]~$\simeq 4.53~\mu_{\rm B}$, and a high-$T$ CW temperature $\theta_{\rm CW}^{\rm HT} \simeq -62.6$~K. This value of $\mu_{\rm eff}^{\rm HT}$ is in good agreement with the expected value of $4.54~\mu_{\rm B}$ for Yb$^{3+}$ ($J = 7/2$, $g=8/7$) in the $^{4}f_{13}$ configuration.


$1/\chi$ is found to change the slope below 80~K and displays a perfect linear behaviour [inset of Fig.~\ref{Fig2}(a)] after substraction of the Van-Vleck susceptibility $\chi_{\rm VV} \simeq 3.32 \times 10^{-3}$~cm$^3$/mol, obtained from the $M$ vs $H$ analysis, discussed later. A CW fit below 50~K yields $\mu_{\rm eff} \simeq 3.2~\mu_{\rm B}$ and $\theta_{\rm CW} \simeq -0.8$~K.
This experimental $\mu_{\rm eff}$ value is reminiscent of an effective spin $J_{\rm eff} = 1/2$ with an average $g \simeq 3.6$. Further, this value of $g$ is significantly larger than the free-electron $g$-value of 2.0 due to the spin-orbit coupling and is consistent with the one extracted from our ESR experiments ($g \simeq 3.4$)~\cite{supplementary}.
Typically, at low temperatures, the magnetic behavior of Yb$^{3+}$-based compounds is governed by the Kramers doublets ($m_{J} = \pm \frac {1}{2}$) triggered by the CEF. In this case, the low-lying ground-state is described by an effective pseudo-spin $J_{\rm eff} = 1/2$ Hamiltonian and the higher-lying levels produce a sizable Van-Vleck contribution ($\chi_{\rm VV}$) to $\chi$~\cite{Li035107,Li167203,Ranjith115804}.
Indeed, several Yb$^{3+}$-based compounds have witnessed $J_{\rm eff} = 1/2$ ground state~\cite{Ranjith180401,Bordelon1058,*Ding144432,Guo094404}. The negative sign of $\theta_{\rm CW}$ indicates dominant AFM intra-plane coupling. Taking the number of NN spins $z = 6$ for a TLAF and the experimental value of $\theta_{\rm CW}$, the NN AFM exchange coupling [$\theta_{\rm CW} = -zJS(S+1)/3k_{\rm B}$] is estimated to be $J/k_{\rm B} \simeq 0.53$~K.

In order to establish the correlated behaviour at low temperatures, the effective magnetic moment {$\mu_{\rm eff}$~[$=\{(3k_B/N_{A}\mu_{\rm B}^{2})\chi T\}^{1/2} = 2.8284\sqrt{\chi T}$}] vs $T$ in different applied fields is plotted in Fig.~\ref{Fig2}(b). At high-$T$s, $\mu_{\rm eff}$ for all the fields approaches the free ion value of $\sim 4.53~\mu_{\rm B}$ and then falls to a plateau of $\sim 3.3$~$\mu_{\rm B}$ in the $T$-range $4 - 20$~K corresponding to $J_{\rm eff} = 1/2$. At low temperatures ($T < 4$~K), the curves deviate significantly from the plateau value. In low fields ($H < 1$~T) $\mu_{\rm eff}(T)$ tends to show an upward curvature while in high fields ($H \geq 1$~T) it shows a downward curvature. This downward curvature in high-fields is a clear testimony of the development of AFM correlation at low temperatures~\cite{Sibille097202}.

Figure~\ref{Fig2}(c) presents the $M$ vs $H$ curves measured at various temperatures in the low-$T$ regime. At $T=0.4$~K, $M$ saturates at around $H_{\rm S} \simeq 0.7$~T and then increases weakly for higher fields due to the Van-Vleck contribution. The value of $H_{\rm S}$ exactly reproduces the intra-plane exchange coupling $J/k_{\rm B}=g \mu_{\rm B} H_{\rm S}/4.5 k_{\rm B} \simeq 0.53$~K, taking $g \simeq 3.4$. This indicates that the inter-layer and/or second NN interactions are negligible and do not contribute to $H_{\rm S}$~\cite{Ranjith180401}. The slope of the linear fit for $H > 3$~T results $\chi_{\rm VV} \simeq 3.32 \times 10^{-3}$~cm$^3$/mol and the $y$-intercept corresponds to the saturation magnetization $M_{\rm sat} \simeq 1.6~\mu_{\rm B}$. This value of $M_{\rm sat}$ provides $g \simeq 3.2$ for $J_{\rm eff} = 1/2$ ($M_{\rm sat} = gJ_{\rm eff}~\mu_{\rm B}$) which is very close to our ESR value ($g=3.4$)~\cite{Ranjith180401} (see SM~\cite{supplementary}).
In the absence of exchange interaction, i.e. in the paramagnetic (PM) state, a magnetic isotherm can be modeled by $M(H) = fgJ_{\rm eff}N_{A}\mu_{\rm B}B_{J_{\rm eff}}(H)$, where $f$ is the molar fraction of free spins and $B_{J_{\rm eff}}(H)$ is the Brillouin function for $J_{\rm eff} = 1/2$~\cite{Kittelc2005}.
Figure~\ref{Fig2}(c) presents the isotherm fits at different temperatures, fixing $g = 3.4$. The value of $f$ is estimated to be about $\sim 0.97$ for all the temperatures. For $T > 1.8$~K, the isotherms are well fitted by the Brillouin function, which comprehends the uncorrelated spins. However, for $T \leq 1.8$~K, the fit deviates significantly from the experimental data reflecting the growth of AFM correlation. In order to visualize this striking feature, we scaled the $T$-axis by dividing $H$ and plotted in Fig.~\ref{Fig2}(d). Above 1.8~K, all the curves collapse on to a single curve, implying the PM nature of the spins while below 1.8~K, the curves progressively deviate from the scaling, ascertaining the emergence of AFM correlation.

\begin{figure}
	\includegraphics[width=\columnwidth]{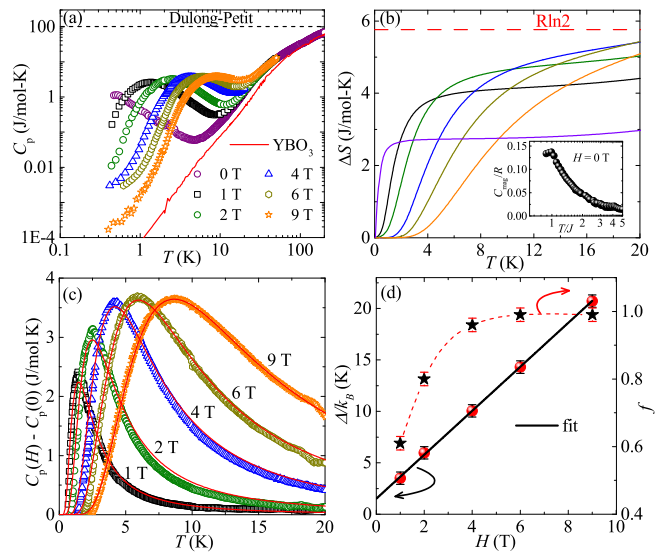}
	\caption{\label{Fig3} (a) $C_{\rm p}$ vs $T$ for YbBO$_3$ measured in different applied magnetic fields. The solid line represents $C_{\rm p}(T)$ of non-magnetic analog YBO$_3$. The horizontal dashed line guides the Dulong-Petit value. (b) Entropy change ($\Delta S$) vs $T$ for different magnetic fields. Inset: $C_{\rm mag}/R$ vs $T/J$ in zero field. (c) Schottky contribution [$C_{\rm p}(H)-C_{\rm p}(0)$] vs $T$ along with the fit using Eq.~\eqref{Eq4}. (d) $\Delta/k_{\rm B}$ and $f$ vs $H$ in the right and left $y$-axes, respectively. The solid line represents the straight line fit to $\Delta/k_{\rm B}(H)$.}
\end{figure}
Heat capacity ($C_{\rm p}$) measured as a function of temperature down to 0.4~K in different magnetic fields is shown in Fig.~\ref{Fig3}(a). At high temperatures, $C_{\rm p}(T)$ is completely dominated by phononic contribution ($C_{\rm ph}$).
Below about 4~K, $C_{\rm p}$ in zero-field increases towards low temperatures as the magnetic correlation sets in and then levels off below $\sim 0.5$~K, similar to NaYb(O,S)$_2$~\cite{Ranjith180401,Baenitz220409}. Absence of any sharp peak rules out the onset of magnetic LRO down to 0.4~K in zero-field. An external magnetic field suppresses the magnetic correlation and gives rise to a broad anomaly which moves towards higher temperatures with field. This broad feature portrays the Schottky anomaly due to the Zeeman splitting of the ground state Kramers doublet.
The magnetic heat capacity ($C_{\rm mag}$) is obtained by subtracting the heat capacity ($C_{\rm ph}$) of the non-magnetic analogue YBO$_3$ from the total $C_{\rm p}$ of YbBO$_3$. The change in magnetic entropy ($\Delta S$) calculated by integrating $C_{\rm mag}/T$ with respect to $T$ 
is plotted in Fig.~\ref{Fig3}(b) for different magnetic fields.
In zero field, $\Delta S$ attains a saturation value of $\sim 2.76$~J/mol-K which is almost 50\% of $R\ln2$, expected for $J_{\rm eff} = 1/2$. This suggests that the remaining 50\% entropy is released below 0.4~K due to the persistence of strong magnetic correlation. The saturation value of $\Delta S$ increases with $H$ and is almost recovered to $R\ln2$ above 2~T ($> H_{\rm S}$).

To probe the Schottky contribution, the zero field data $C_{\rm p}(T,H=0)$ are subtracted from the high field data $C_{\rm p}(T,H)$ [i.e. $C_{\rm Sch}(T,H) = C_{\rm p}(T,H) - C_{\rm p}(T,H = 0)$]. Figure~\ref{Fig3}(c) shows the fit of $C_{\rm p}(T,H) - C_{\rm p}(T,H = 0)$ data using the two level Schottky function~\cite{Kittelc2005}
\begin{equation}
	C_{\rm Sch}(T) = fR\left(\frac{\Delta}{k_{\rm B}T}\right)^{2} \frac{e^{(\Delta/k_{\rm B}T)}}{\left[1+e^{(\Delta/k_{\rm B}T)}\right]^2}.
	\label{Eq4}
\end{equation}
Here, $\Delta/k_{\rm B}$ is the crystal field gap between the ground state and the first excited state doublets and $R$ is the gas constant. The estimated $f$ and $\Delta/k_{\rm B}$ values are plotted as a function of $H$ in the right and left $y$-axes, respectively in Fig.~\ref{Fig3}(d). $f$ increases with $H$ and then saturates to a value of about $\sim 1$ for $H>2$~T. This confirms that magnetic field splits the energy levels and excites the free Yb$^{3+}$ spins to the higher energy levels. For $H < H_{\rm S}$, a fraction of spins are correlated which is reduced with increasing field and above $H_{\rm S}$, all the free spins are excited. This explains why $\Delta S$ increases with $H$ and then attains a saturation value $R ln 2$ for $H > H_{\rm S}$~\cite{BagL220403}.
Similarly, the peak maximum of the $C_{\rm p}(H) - C_{\rm p}(0)$ curves remains unchanged for $H>2$~T, suggesting that $\sim 100$\% spins become free in higher fields [see Fig.~\ref{Fig3}(c)]. 
In Fig.~\ref{Fig3}(d), $\Delta/k_{\rm B}$ increases linearly with $H$ and a straight line fit results the zero-field energy-gap $\Delta/k_{\rm B}(0) \simeq 1.5$~K which possibly indicates an intrinsic field in the system~\cite{Kundu117206}. From the value of $\Delta/k_{\rm B} \simeq 20.6$~K at 9~T, one can estimate the $g$-value as $\Delta/k_B = g\mu_{\rm B}H/k_B$ that yields $g \simeq 3.3$, consistent with the ESR value.

Furthermore, $C_{\rm mag}(T)$ in the inset of Fig.~\ref{Fig3}(b) shows a plateau at $\sim 0.5$~K with a maximum value $C_{\rm mag}^{\rm max}/R \simeq 0.137$. For a frustrated spin-$1/2$ TLAF one expects such a broad maximum at $T/J \simeq 0.84$ but with a reduced value $C_{\rm mag}/R \simeq 0.223$~\cite{Bernu134409} compared to a non-frustrated 2D antiferromagnet (0.44)~\cite{Makivic3562,*Kim2705}. Though, our maximum appears nearly at the anticipated position ($T/J \simeq 0.9$) but the absolute value is significantly lower than that expected for a TLAF. Similar peak type feature in YbMgGaO$_4$ and NaYbS$_2$ is ascribed to the disorder induced and disorder free QSLs, respectively~\cite{Li16419,Baenitz220409}.
In order to examine this peculiar feature, $\mu$SR experiments are carried out in zero field (ZF).

\begin{figure}
	\includegraphics[width=\columnwidth]{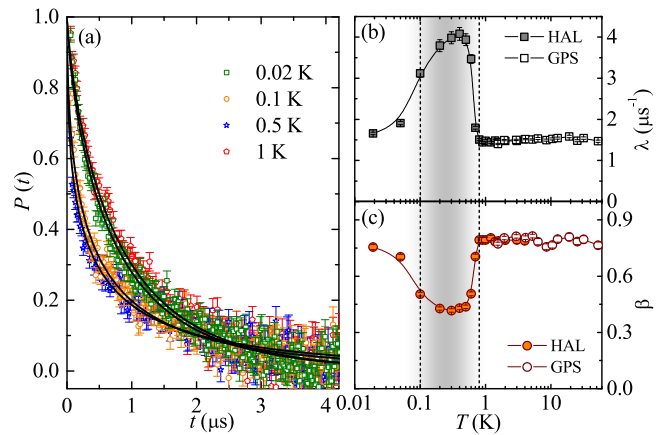}
	\caption{(a) Muon decay asymmetry as a function of time, with solid lines representing the fit using stretched exponential function. Temperature dependence of (b) depolarization rate ($\lambda$) and (c) stretching parameter ($\beta$). The shaded portion highlights the fluctuating regime.
		\label{Fig4}}
\end{figure}

Muon spin is susceptible to extremely small magnetic field ($\sim 10^{-5}$~T) and because of its much shorter time window (10~ns–15~$\mu$s), $\mu$SR is an excellent probe in tracing the dynamics of slowly fluctuating magnets.
In general, the ZF $\mu^+$ spin depolarization rate ($\lambda$) is related to spin-spin correlation function $\lambda_{\rm ZF} = \gamma_{\mu}^{2}\int_{0}^{\infty} \langle B_{\rm loc}^{\perp}(t)\cdot B_{\rm loc}^{\perp}(0) \rangle \,dt \propto S_{\omega \to 0}^{\perp}$, where $S_{\omega \to 0}^{\perp} = \int_{0}^{\infty} \langle s_{i}^{\perp}(t)\cdot s_{i}^{\perp}(0) \rangle \,dt$ is the static spin structure factor~\cite{Avella2013,Li097201}. Hence, by measuring muon asymmetry as a function of temperature one can comprehend the correlated behaviour of a spin system.
Figure~\ref{Fig4}(a) depicts the muon asymmetry curves at different temperatures down to 20~mK which are fitted well by a stretched exponential function, $P(t) = P(0) e^{-(\lambda t)^\beta}$.
Here, $P(0)$ ($\sim 0.118$ and 0.213 for HAL and GPS, respectively) is the initial asymmetry.
The introduction of stretching parameter $\beta$ suggests that there is a distribution of the relaxation rates. The obtained $\lambda$ and $\beta$ as a function of $T$ are summarized in Fig.~\ref{Fig4}(b) and (c), respectively.

At high temperatures, the magnetic moments fluctuate very fast and only mildly depolarize the implanted muon before it decays and the dominant depolarization channel is due to nuclear spins. This gives rise to a temperature independent $\lambda$, as we observed in Fig.~\ref{Fig4}(b). Similarly, $\beta$ remains temperature independent in high temperatures and it's value is slightly below 1, indicating a distribution likely due to weak disorder in the O-site (muon stopping site) or a few muon sites. As the temperature is lowered below $\sim 0.9$~K which is close to $\theta_{\rm CW}$, $\beta$ drops and $\lambda$ increases rapidly exhibiting a broad minima and maxima, respectively centered at $T^* \simeq 0.4$~K. The rapid increase in $\lambda$ suggest the slowing down of the fluctuating moments as the correlation sets in and replicates the behaviour of $C_{\rm p}(T)$. Upon further lowering the temperature, the values of $\beta$ and $\lambda$ recover back to their initial values below $\sim 0.1$~K, clearly suggesting an exotic dynamical regime (0.1~K~$\leq T \leq 0.9$~K) with slow fluctuations, marked by shading in Fig.~\ref{Fig4}(b) and (c)~\cite{Lee224420}.
In the PM regime ($T > 1$~K) where $\lambda_{\rm ZF}$ is temperature independent, the spin fluctuation rate ($\nu$) can be calculated as $\nu = \sqrt{z}JS/\hbar \sim 8.542 \times 10^{10}$~Hz taking $z = 6$, $S=1/2$, and $J/k_{\rm B} \simeq 0.53$~K~\cite{Uemura3306,Lee024413}. In the fast fluctuation limit, using the relation $\lambda_{\rm ZF} = 2\Delta_{\mu}^{2}/\nu$, the distribution width of the local magnetic fields is estimated to be $\Delta_{\mu} \simeq 254$~MHz $\ll \nu$ considering $\lambda_{\rm ZF}~(T > 1~{\rm K}) \simeq 1.5~\mu{\rm s}^{-1}$.

In case of a conventional magnetic LRO, the muon decay asymmetry should exhibit an oscillating signal while for a spin-glass (SG) transition, it should feature a non-oscillating and undamped $1/3$ tail for the polycrystalline samples~\cite{Li097201}. Further, for a SG transition, $\beta$ is expected to drop to $\sim 0.33$ as $T$ tends to $T_{\rm g}$ and $\lambda$ to show a peak or divergence~\cite{Keren1386,Onishi184412,Campbell1291}. Absence of any such signatures in Fig.~\ref{Fig4} rules out both a static magnetic LRO and a SG transition.
We also do not observe any saturation of the increased $\lambda$ at low temperatures, in contrast to that reported for many QSL candidates due to persistent spin fluctuations~\cite{Mendels077204,Li097201,Lee024413}.
Additionally, though, $\beta$ value drops to $\sim 0.4$ at $T^*$ but it recovers at low temperatures without settling to a particular value as empirically observed in QSL systems. These observations exclude the QSL scenario and imply that the slowly fluctuating regime is restricted to $0.182 \leq T/J \leq 1.63$ which could be due to short-range correlations.

The observed peculiar behaviour prompted us to compare the structural aspects of the Yb$^{3+}$ based TLAFs. The ratio of interlayer to intralayer spacing $d_{\rm inter}/d_{\rm intra} \simeq 1.17$ in YbBO$_3$ is smaller compared to NaYb(O,S,Se)$_2$ ($\sim 1.64 - 1.71$) and YbGaMgO$_4$ ($\sim 2.4$) but almost equal to NaBaYb(BO$_3$)$_2$ ($\sim 1.1$)~\cite{Guo10670}. In NaYb(O,S,Se)$_2$, the Yb$^{3+}$ layers follow $ABC$ stacking. This staggered arrangement generates inter-layer frustration which possibly forbids magnetic LRO in zero-field and foster QSL state~\cite{Ranjith180401,Ranjith224417,Baenitz220409,Bordelon224427}. Similarly, for YbGaMgO$_4$, large $d_{\rm inter}$ leads to pronounced two-dimensionality, $ABC$ stacking produces inter-layer frustration, and non-magnetic Mg/Ga site mixing results bond-randomness~\cite{Kimchi031028}. Consequently, the dominant effect from randomness along with magnetic frustration is responsible for curbing of magnetic LRO and establishing QSL-like behaviour in YbGaMgO$_4$~\cite{Kimchi031028,Li107202,Li167203}. On the other hand, NaBaYb(BO$_3$)$_2$ which has $ABC$ stacking, does not possess any structural disorder and undergoes a magnetic LRO at $T_{\rm N} \simeq 0.4$~K~\cite{Guo094404,Guo094404,Pan104412}. 
Despite the same $d_{\rm inter}/d_{\rm intra}$ ratio as that of NaBaYb(BO$_3$)$_2$ and with $AAA$ stacking, YbBO$_3$ manifests no magnetic LRO and exhibits a unconventional fluctuating regime.

Interestingly, the behaviour of $\lambda(T)$ resembles with that of the $S=3/2$ TLAFs $\alpha$-HCrO$_2$ and NaCrO$_2$~\cite{Somesh104422,Olariu167203}. In these compounds, there exists a broad fluctuating regime with slow dynamics at low temperatures followed by the formation of a static order at very low temperatures. The $\lambda$ vs $T/J$ plot divulges the broad maximum centered at the same position ($T/J \simeq 0.6-0.7$) for both compounds, establishing that the broad dynamical regime is universal to TLAFs. These chromates belong to the delafossite family where the interlayer frustration due to $ABC$-stacking of the magnetic layers was believed to be the main reason for this unconventional dynamics. For YbBO$_3$, $C_{\rm p}(T)$ shows a very broad peak and the center position of the broad maxima in $\lambda$ corresponds to $T/J \simeq 0.7$ with $J/k_{\rm B} \simeq 0.53$~K, quite similar to the chromates. However, unlike the delafossites, the Yb$^{3+}$ layers in YbBO$_3$ are arranged in a $AAA$-stacking sequence which is expected to produce a non-frustrated interlayer geometry and should favour magnetic LRO at a finite temperature depending on the strength of the inter-layer coupling~\cite{Bordelon224427}.

A closer inspection of the crystal structure revealed that the Yb$^{3+}$ ions are eight-fold coordinated with oxygen [six O(1) and six O(2)] atoms forming YbO$_8$ polyhedra~\cite{Chadeyron261}. The polyhedra of two adjacent layers are connected only via partially occupied O(2) and B sites which provide the inter-layer interaction path. This positional disorder of O(2) and B atoms adversely affects the local environment of Yb$^{3+}$ and may induce bond randomness, similar to YbMgGaO$_4$. Secondly, it also weakens the inter-layer interaction which is responsible for the onset of three-dimensional (3D) magnetic LRO. As a consequence, the 3D magnetic LRO is prevented, rendering the spin-lattice an ideal 2D system. Thus, the absence of magnetic LRO can be accredited to the non-magnetic site disorder and the intermediate fluctuating regime with slow dynamics could be a universal feature of isotropic TLAFs arising due to strong intra-layer frustration.

In summary, we performed a comprehensive study of the static and dynamic properties of a new quantum magnet YbBO$_3$ possessing isotropic Yb$^{3+}$ triangular lattice. The low temperature properties are described well by $J_{\rm eff}=1/2$ Kramers doublets of the Yb$^{3+}$ ions.
Neither magnetic LRO nor spin freezing are witnessed down to 20~mK.
setting the lower limit of the frustration parameter $f (= |\theta_{\rm CW}|/T_{\rm N}) \simeq 40$, a factual hallmark of a highly frustrated magnet. The analysis of $\chi(T)$ and magnetic isotherms suggests the emergence of AFM correlation below $\sim 0.9$~K. The $\mu$SR depolarization rate together with heat capacity data demonstrate an extended fluctuating regime with slow dynamics at low temperatures. This unusual dynamics and the absence of magnetic LRO are ascribed to the effects of intra-layer frustration of a perfect 2D TLAF and positional disorder, respectively. With these features YbBO$_3$ stands out as an exception among the Yb$^{3+}$-based TLAFs.

\section{Acknowledgments}
We would like to acknowledge SERB, India for financial support bearing sanction Grant No. CRG/2019/000960. Part of this work is based on experiments performed at the Swiss Muon Source S$\mu$S, Paul Scherrer Institute, Villigen, Switzerland. The project has received funding from the European Union’s Horizon 2020 research and innovation programme under the Marie Skłodowska-Curie grant agreement No 884104 (PSI-FELLOW-III-3i). We thank Robert Scheuermann for his support with the HAL instrument.


\begin{thebibliography}{62}%
	\makeatletter
	\providecommand \@ifxundefined [1]{%
		\@ifx{#1\undefined}
	}%
	\providecommand \@ifnum [1]{%
		\ifnum #1\expandafter \@firstoftwo
		\else \expandafter \@secondoftwo
		\fi
	}%
	\providecommand \@ifx [1]{%
		\ifx #1\expandafter \@firstoftwo
		\else \expandafter \@secondoftwo
		\fi
	}%
	\providecommand \natexlab [1]{#1}%
	\providecommand \enquote  [1]{``#1''}%
	\providecommand \bibnamefont  [1]{#1}%
	\providecommand \bibfnamefont [1]{#1}%
	\providecommand \citenamefont [1]{#1}%
	\providecommand \href@noop [0]{\@secondoftwo}%
	\providecommand \href [0]{\begingroup \@sanitize@url \@href}%
	\providecommand \@href[1]{\@@startlink{#1}\@@href}%
	\providecommand \@@href[1]{\endgroup#1\@@endlink}%
	\providecommand \@sanitize@url [0]{\catcode `\\12\catcode `\$12\catcode
		`\&12\catcode `\#12\catcode `\^12\catcode `\_12\catcode `\%12\relax}%
	\providecommand \@@startlink[1]{}%
	\providecommand \@@endlink[0]{}%
	\providecommand \url  [0]{\begingroup\@sanitize@url \@url }%
	\providecommand \@url [1]{\endgroup\@href {#1}{\urlprefix }}%
	\providecommand \urlprefix  [0]{URL }%
	\providecommand \Eprint [0]{\href }%
	\providecommand \doibase [0]{https://doi.org/}%
	\providecommand \selectlanguage [0]{\@gobble}%
	\providecommand \bibinfo  [0]{\@secondoftwo}%
	\providecommand \bibfield  [0]{\@secondoftwo}%
	\providecommand \translation [1]{[#1]}%
	\providecommand \BibitemOpen [0]{}%
	\providecommand \bibitemStop [0]{}%
	\providecommand \bibitemNoStop [0]{.\EOS\space}%
	\providecommand \EOS [0]{\spacefactor3000\relax}%
	\providecommand \BibitemShut  [1]{\csname bibitem#1\endcsname}%
	\let\auto@bib@innerbib\@empty
	\bibitem [{\citenamefont {Balents}(2010)}]{Balents199}%
	\BibitemOpen
	\bibfield  {author} {\bibinfo {author} {\bibfnamefont {L.}~\bibnamefont
			{Balents}},\ }\href {https://doi.org/10.1038/nature08917} {\bibfield
		{journal} {\bibinfo  {journal} {Nature}\ }\textbf {\bibinfo {volume} {464}},\
		\bibinfo {pages} {199} (\bibinfo {year} {2010})}\BibitemShut {NoStop}%
	\bibitem [{\citenamefont {Broholm}\ \emph {et~al.}(2020)\citenamefont
		{Broholm}, \citenamefont {Cava}, \citenamefont {Kivelson}, \citenamefont
		{Nocera}, \citenamefont {Norman},\ and\ \citenamefont
		{Senthil}}]{Broholmeaay0668}%
	\BibitemOpen
	\bibfield  {author} {\bibinfo {author} {\bibfnamefont {C.}~\bibnamefont
			{Broholm}}, \bibinfo {author} {\bibfnamefont {R.~J.}\ \bibnamefont {Cava}},
		\bibinfo {author} {\bibfnamefont {S.~A.}\ \bibnamefont {Kivelson}}, \bibinfo
		{author} {\bibfnamefont {D.~G.}\ \bibnamefont {Nocera}}, \bibinfo {author}
		{\bibfnamefont {M.~R.}\ \bibnamefont {Norman}},\ and\ \bibinfo {author}
		{\bibfnamefont {T.}~\bibnamefont {Senthil}},\ }\href
	{https://doi.org/10.1126/science.aay0668} {\bibfield  {journal} {\bibinfo
			{journal} {Science}\ }\textbf {\bibinfo {volume} {367}},\ \bibinfo {pages}
		{eaay0668} (\bibinfo {year} {2020})}\BibitemShut {NoStop}%
	\bibitem [{\citenamefont {Anderson}(1973)}]{Anderson153}%
	\BibitemOpen
	\bibfield  {author} {\bibinfo {author} {\bibfnamefont {P.}~\bibnamefont
			{Anderson}},\ }\href
	{https://doi.org/https://doi.org/10.1016/0025-5408(73)90167-0} {\bibfield
		{journal} {\bibinfo  {journal} {Mater. Res. Bull}\ }\textbf {\bibinfo
			{volume} {8}},\ \bibinfo {pages} {153 } (\bibinfo {year} {1973})}\BibitemShut
	{NoStop}%
	\bibitem [{\citenamefont {Singh}\ and\ \citenamefont {Huse}(1992)}]{Singh1766}%
	\BibitemOpen
	\bibfield  {author} {\bibinfo {author} {\bibfnamefont {R.~R.~P.}\
			\bibnamefont {Singh}}\ and\ \bibinfo {author} {\bibfnamefont {D.~A.}\
			\bibnamefont {Huse}},\ }\href {https://doi.org/10.1103/PhysRevLett.68.1766}
	{\bibfield  {journal} {\bibinfo  {journal} {Phys. Rev. Lett.}\ }\textbf
		{\bibinfo {volume} {68}},\ \bibinfo {pages} {1766} (\bibinfo {year}
		{1992})}\BibitemShut {NoStop}%
	\bibitem [{\citenamefont {Capriotti}\ \emph {et~al.}(1999)\citenamefont
		{Capriotti}, \citenamefont {Trumper},\ and\ \citenamefont
		{Sorella}}]{Capriotti3899}%
	\BibitemOpen
	\bibfield  {author} {\bibinfo {author} {\bibfnamefont {L.}~\bibnamefont
			{Capriotti}}, \bibinfo {author} {\bibfnamefont {A.~E.}\ \bibnamefont
			{Trumper}},\ and\ \bibinfo {author} {\bibfnamefont {S.}~\bibnamefont
			{Sorella}},\ }\href {https://doi.org/10.1103/PhysRevLett.82.3899} {\bibfield
		{journal} {\bibinfo  {journal} {Phys. Rev. Lett.}\ }\textbf {\bibinfo
			{volume} {82}},\ \bibinfo {pages} {3899} (\bibinfo {year}
		{1999})}\BibitemShut {NoStop}%
	\bibitem [{\citenamefont {Watanabe}\ \emph {et~al.}(2014)\citenamefont
		{Watanabe}, \citenamefont {Kawamura}, \citenamefont {Nakano},\ and\
		\citenamefont {Sakai}}]{Watanabe034714}%
	\BibitemOpen
	\bibfield  {author} {\bibinfo {author} {\bibfnamefont {K.}~\bibnamefont
			{Watanabe}}, \bibinfo {author} {\bibfnamefont {H.}~\bibnamefont {Kawamura}},
		\bibinfo {author} {\bibfnamefont {H.}~\bibnamefont {Nakano}},\ and\ \bibinfo
		{author} {\bibfnamefont {T.}~\bibnamefont {Sakai}},\ }\href
	{https://doi.org/10.7566/JPSJ.83.034714} {\bibfield  {journal} {\bibinfo
			{journal} {J. Phys. Soc. Jpn}\ }\textbf {\bibinfo {volume} {83}},\ \bibinfo
		{pages} {034714} (\bibinfo {year} {2014})}\BibitemShut {NoStop}%
	\bibitem [{\citenamefont {Kimchi}\ \emph {et~al.}(2018)\citenamefont {Kimchi},
		\citenamefont {Nahum},\ and\ \citenamefont {Senthil}}]{Kimchi031028}%
	\BibitemOpen
	\bibfield  {author} {\bibinfo {author} {\bibfnamefont {I.}~\bibnamefont
			{Kimchi}}, \bibinfo {author} {\bibfnamefont {A.}~\bibnamefont {Nahum}},\ and\
		\bibinfo {author} {\bibfnamefont {T.}~\bibnamefont {Senthil}},\ }\href
	{https://doi.org/10.1103/PhysRevX.8.031028} {\bibfield  {journal} {\bibinfo
			{journal} {Phys. Rev. X}\ }\textbf {\bibinfo {volume} {8}},\ \bibinfo {pages}
		{031028} (\bibinfo {year} {2018})}\BibitemShut {NoStop}%
	\bibitem [{\citenamefont {Li}\ \emph {et~al.}(2015{\natexlab{a}})\citenamefont
		{Li}, \citenamefont {Bishop},\ and\ \citenamefont {Campbell}}]{Li014426}%
	\BibitemOpen
	\bibfield  {author} {\bibinfo {author} {\bibfnamefont {P.~H.~Y.}\
			\bibnamefont {Li}}, \bibinfo {author} {\bibfnamefont {R.~F.}\ \bibnamefont
			{Bishop}},\ and\ \bibinfo {author} {\bibfnamefont {C.~E.}\ \bibnamefont
			{Campbell}},\ }\href {https://doi.org/10.1103/PhysRevB.91.014426} {\bibfield
		{journal} {\bibinfo  {journal} {Phys. Rev. B}\ }\textbf {\bibinfo {volume}
			{91}},\ \bibinfo {pages} {014426} (\bibinfo {year}
		{2015}{\natexlab{a}})}\BibitemShut {NoStop}%
	\bibitem [{\citenamefont {Maksimov}\ \emph {et~al.}(2019)\citenamefont
		{Maksimov}, \citenamefont {Zhu}, \citenamefont {White},\ and\ \citenamefont
		{Chernyshev}}]{Maksimov021017}%
	\BibitemOpen
	\bibfield  {author} {\bibinfo {author} {\bibfnamefont {P.~A.}\ \bibnamefont
			{Maksimov}}, \bibinfo {author} {\bibfnamefont {Z.}~\bibnamefont {Zhu}},
		\bibinfo {author} {\bibfnamefont {S.~R.}\ \bibnamefont {White}},\ and\
		\bibinfo {author} {\bibfnamefont {A.~L.}\ \bibnamefont {Chernyshev}},\ }\href
	{https://doi.org/10.1103/PhysRevX.9.021017} {\bibfield  {journal} {\bibinfo
			{journal} {Phys. Rev. X}\ }\textbf {\bibinfo {volume} {9}},\ \bibinfo {pages}
		{021017} (\bibinfo {year} {2019})}\BibitemShut {NoStop}%
	\bibitem [{\citenamefont {Somesh}\ \emph {et~al.}(2021)\citenamefont {Somesh},
		\citenamefont {Furukawa}, \citenamefont {Simutis}, \citenamefont {Bert},
		\citenamefont {Prinz-Zwick}, \citenamefont {B\"uttgen}, \citenamefont
		{Zorko}, \citenamefont {Tsirlin}, \citenamefont {Mendels},\ and\
		\citenamefont {Nath}}]{Somesh104422}%
	\BibitemOpen
	\bibfield  {author} {\bibinfo {author} {\bibfnamefont {K.}~\bibnamefont
			{Somesh}}, \bibinfo {author} {\bibfnamefont {Y.}~\bibnamefont {Furukawa}},
		\bibinfo {author} {\bibfnamefont {G.}~\bibnamefont {Simutis}}, \bibinfo
		{author} {\bibfnamefont {F.}~\bibnamefont {Bert}}, \bibinfo {author}
		{\bibfnamefont {M.}~\bibnamefont {Prinz-Zwick}}, \bibinfo {author}
		{\bibfnamefont {N.}~\bibnamefont {B\"uttgen}}, \bibinfo {author}
		{\bibfnamefont {A.}~\bibnamefont {Zorko}}, \bibinfo {author} {\bibfnamefont
			{A.~A.}\ \bibnamefont {Tsirlin}}, \bibinfo {author} {\bibfnamefont
			{P.}~\bibnamefont {Mendels}},\ and\ \bibinfo {author} {\bibfnamefont
			{R.}~\bibnamefont {Nath}},\ }\href
	{https://doi.org/10.1103/PhysRevB.104.104422} {\bibfield  {journal} {\bibinfo
			{journal} {Phys. Rev. B}\ }\textbf {\bibinfo {volume} {104}},\ \bibinfo
		{pages} {104422} (\bibinfo {year} {2021})}\BibitemShut {NoStop}%
	\bibitem [{\citenamefont {Olariu}\ \emph {et~al.}(2006)\citenamefont {Olariu},
		\citenamefont {Mendels}, \citenamefont {Bert}, \citenamefont {Ueland},
		\citenamefont {Schiffer}, \citenamefont {Berger},\ and\ \citenamefont
		{Cava}}]{Olariu167203}%
	\BibitemOpen
	\bibfield  {author} {\bibinfo {author} {\bibfnamefont {A.}~\bibnamefont
			{Olariu}}, \bibinfo {author} {\bibfnamefont {P.}~\bibnamefont {Mendels}},
		\bibinfo {author} {\bibfnamefont {F.}~\bibnamefont {Bert}}, \bibinfo {author}
		{\bibfnamefont {B.~G.}\ \bibnamefont {Ueland}}, \bibinfo {author}
		{\bibfnamefont {P.}~\bibnamefont {Schiffer}}, \bibinfo {author}
		{\bibfnamefont {R.~F.}\ \bibnamefont {Berger}},\ and\ \bibinfo {author}
		{\bibfnamefont {R.~J.}\ \bibnamefont {Cava}},\ }\href
	{https://doi.org/10.1103/PhysRevLett.97.167203} {\bibfield  {journal}
		{\bibinfo  {journal} {Phys. Rev. Lett.}\ }\textbf {\bibinfo {volume} {97}},\
		\bibinfo {pages} {167203} (\bibinfo {year} {2006})}\BibitemShut {NoStop}%
	\bibitem [{\citenamefont {Kaneko}\ \emph {et~al.}(2014)\citenamefont {Kaneko},
		\citenamefont {Morita},\ and\ \citenamefont {Imada}}]{Kaneko093707}%
	\BibitemOpen
	\bibfield  {author} {\bibinfo {author} {\bibfnamefont {R.}~\bibnamefont
			{Kaneko}}, \bibinfo {author} {\bibfnamefont {S.}~\bibnamefont {Morita}},\
		and\ \bibinfo {author} {\bibfnamefont {M.}~\bibnamefont {Imada}},\ }\href
	{https://doi.org/10.7566/JPSJ.83.093707} {\bibfield  {journal} {\bibinfo
			{journal} {J. Phys. Soc. Jpn.}\ }\textbf {\bibinfo {volume} {83}},\ \bibinfo
		{pages} {093707} (\bibinfo {year} {2014})}\BibitemShut {NoStop}%
	\bibitem [{\citenamefont {Li}\ \emph {et~al.}(2020)\citenamefont {Li},
		\citenamefont {Gegenwart},\ and\ \citenamefont {Tsirlin}}]{Li224004}%
	\BibitemOpen
	\bibfield  {author} {\bibinfo {author} {\bibfnamefont {Y.}~\bibnamefont
			{Li}}, \bibinfo {author} {\bibfnamefont {P.}~\bibnamefont {Gegenwart}},\ and\
		\bibinfo {author} {\bibfnamefont {A.~A.}\ \bibnamefont {Tsirlin}},\ }\href
	{https://doi.org/10.1088/1361-648X/ab724e} {\bibfield  {journal} {\bibinfo
			{journal} {J. Phys.: Condens. Matter}\ }\textbf {\bibinfo {volume} {32}},\
		\bibinfo {pages} {224004} (\bibinfo {year} {2020})}\BibitemShut {NoStop}%
	\bibitem [{\citenamefont {Li}\ \emph {et~al.}(2016{\natexlab{a}})\citenamefont
		{Li}, \citenamefont {Wang},\ and\ \citenamefont {Chen}}]{Li035107}%
	\BibitemOpen
	\bibfield  {author} {\bibinfo {author} {\bibfnamefont {Y.~D.}\ \bibnamefont
			{Li}}, \bibinfo {author} {\bibfnamefont {X.}~\bibnamefont {Wang}},\ and\
		\bibinfo {author} {\bibfnamefont {G.}~\bibnamefont {Chen}},\ }\href
	{https://doi.org/10.1103/PhysRevB.94.035107} {\bibfield  {journal} {\bibinfo
			{journal} {Phys. Rev. B}\ }\textbf {\bibinfo {volume} {94}},\ \bibinfo
		{pages} {035107} (\bibinfo {year} {2016}{\natexlab{a}})}\BibitemShut
	{NoStop}%
	\bibitem [{\citenamefont {Furukawa}\ \emph {et~al.}(2015)\citenamefont
		{Furukawa}, \citenamefont {Miyagawa}, \citenamefont {Itou}, \citenamefont
		{Ito}, \citenamefont {Taniguchi}, \citenamefont {Saito}, \citenamefont
		{Iguchi}, \citenamefont {Sasaki},\ and\ \citenamefont
		{Kanoda}}]{Furukawa077001}%
	\BibitemOpen
	\bibfield  {author} {\bibinfo {author} {\bibfnamefont {T.}~\bibnamefont
			{Furukawa}}, \bibinfo {author} {\bibfnamefont {K.}~\bibnamefont {Miyagawa}},
		\bibinfo {author} {\bibfnamefont {T.}~\bibnamefont {Itou}}, \bibinfo {author}
		{\bibfnamefont {M.}~\bibnamefont {Ito}}, \bibinfo {author} {\bibfnamefont
			{H.}~\bibnamefont {Taniguchi}}, \bibinfo {author} {\bibfnamefont
			{M.}~\bibnamefont {Saito}}, \bibinfo {author} {\bibfnamefont
			{S.}~\bibnamefont {Iguchi}}, \bibinfo {author} {\bibfnamefont
			{T.}~\bibnamefont {Sasaki}},\ and\ \bibinfo {author} {\bibfnamefont
			{K.}~\bibnamefont {Kanoda}},\ }\href
	{https://doi.org/10.1103/PhysRevLett.115.077001} {\bibfield  {journal}
		{\bibinfo  {journal} {Phys. Rev. Lett.}\ }\textbf {\bibinfo {volume} {115}},\
		\bibinfo {pages} {077001} (\bibinfo {year} {2015})}\BibitemShut {NoStop}%
	\bibitem [{\citenamefont {Li}\ \emph {et~al.}(2015{\natexlab{b}})\citenamefont
		{Li}, \citenamefont {Chen}, \citenamefont {Tong}, \citenamefont {Pi},
		\citenamefont {Liu}, \citenamefont {Yang}, \citenamefont {Wang},\ and\
		\citenamefont {Zhang}}]{Li167203}%
	\BibitemOpen
	\bibfield  {author} {\bibinfo {author} {\bibfnamefont {Y.}~\bibnamefont
			{Li}}, \bibinfo {author} {\bibfnamefont {G.}~\bibnamefont {Chen}}, \bibinfo
		{author} {\bibfnamefont {W.}~\bibnamefont {Tong}}, \bibinfo {author}
		{\bibfnamefont {L.}~\bibnamefont {Pi}}, \bibinfo {author} {\bibfnamefont
			{J.}~\bibnamefont {Liu}}, \bibinfo {author} {\bibfnamefont {Z.}~\bibnamefont
			{Yang}}, \bibinfo {author} {\bibfnamefont {X.}~\bibnamefont {Wang}},\ and\
		\bibinfo {author} {\bibfnamefont {Q.}~\bibnamefont {Zhang}},\ }\href
	{https://doi.org/10.1103/PhysRevLett.115.167203} {\bibfield  {journal}
		{\bibinfo  {journal} {Phys. Rev. Lett.}\ }\textbf {\bibinfo {volume} {115}},\
		\bibinfo {pages} {167203} (\bibinfo {year} {2015}{\natexlab{b}})}\BibitemShut
	{NoStop}%
	\bibitem [{\citenamefont {Li}\ \emph {et~al.}(2015{\natexlab{c}})\citenamefont
		{Li}, \citenamefont {Liao}, \citenamefont {Zhang}, \citenamefont {Li},
		\citenamefont {Jin}, \citenamefont {Ling}, \citenamefont {Zhang},
		\citenamefont {Zou}, \citenamefont {Pi}, \citenamefont {Yang}, \citenamefont
		{Wang}, \citenamefont {Wu},\ and\ \citenamefont {Zhang}}]{Li16419}%
	\BibitemOpen
	\bibfield  {author} {\bibinfo {author} {\bibfnamefont {Y.}~\bibnamefont
			{Li}}, \bibinfo {author} {\bibfnamefont {H.}~\bibnamefont {Liao}}, \bibinfo
		{author} {\bibfnamefont {Z.}~\bibnamefont {Zhang}}, \bibinfo {author}
		{\bibfnamefont {S.}~\bibnamefont {Li}}, \bibinfo {author} {\bibfnamefont
			{F.}~\bibnamefont {Jin}}, \bibinfo {author} {\bibfnamefont {L.}~\bibnamefont
			{Ling}}, \bibinfo {author} {\bibfnamefont {L.}~\bibnamefont {Zhang}},
		\bibinfo {author} {\bibfnamefont {Y.}~\bibnamefont {Zou}}, \bibinfo {author}
		{\bibfnamefont {L.}~\bibnamefont {Pi}}, \bibinfo {author} {\bibfnamefont
			{Z.}~\bibnamefont {Yang}}, \bibinfo {author} {\bibfnamefont {J.}~\bibnamefont
			{Wang}}, \bibinfo {author} {\bibfnamefont {Z.}~\bibnamefont {Wu}},\ and\
		\bibinfo {author} {\bibfnamefont {Q.}~\bibnamefont {Zhang}},\ }\href
	{https://doi.org/10.1038/srep16419} {\bibfield  {journal} {\bibinfo
			{journal} {Sci. Rep.}\ }\textbf {\bibinfo {volume} {5}},\ \bibinfo {pages}
		{16419} (\bibinfo {year} {2015}{\natexlab{c}})}\BibitemShut {NoStop}%
	\bibitem [{\citenamefont {Li}\ \emph {et~al.}(2017)\citenamefont {Li},
		\citenamefont {Adroja}, \citenamefont {Bewley}, \citenamefont {Voneshen},
		\citenamefont {Tsirlin}, \citenamefont {Gegenwart},\ and\ \citenamefont
		{Zhang}}]{Li107202}%
	\BibitemOpen
	\bibfield  {author} {\bibinfo {author} {\bibfnamefont {Y.}~\bibnamefont
			{Li}}, \bibinfo {author} {\bibfnamefont {D.}~\bibnamefont {Adroja}}, \bibinfo
		{author} {\bibfnamefont {R.~I.}\ \bibnamefont {Bewley}}, \bibinfo {author}
		{\bibfnamefont {D.}~\bibnamefont {Voneshen}}, \bibinfo {author}
		{\bibfnamefont {A.~A.}\ \bibnamefont {Tsirlin}}, \bibinfo {author}
		{\bibfnamefont {P.}~\bibnamefont {Gegenwart}},\ and\ \bibinfo {author}
		{\bibfnamefont {Q.}~\bibnamefont {Zhang}},\ }\href
	{https://doi.org/10.1103/PhysRevLett.118.107202} {\bibfield  {journal}
		{\bibinfo  {journal} {Phys. Rev. Lett.}\ }\textbf {\bibinfo {volume} {118}},\
		\bibinfo {pages} {107202} (\bibinfo {year} {2017})}\BibitemShut {NoStop}%
	\bibitem [{\citenamefont {Zhu}\ \emph {et~al.}(2017)\citenamefont {Zhu},
		\citenamefont {Maksimov}, \citenamefont {White},\ and\ \citenamefont
		{Chernyshev}}]{Zhu157201}%
	\BibitemOpen
	\bibfield  {author} {\bibinfo {author} {\bibfnamefont {Z.}~\bibnamefont
			{Zhu}}, \bibinfo {author} {\bibfnamefont {P.~A.}\ \bibnamefont {Maksimov}},
		\bibinfo {author} {\bibfnamefont {S.~R.}\ \bibnamefont {White}},\ and\
		\bibinfo {author} {\bibfnamefont {A.~L.}\ \bibnamefont {Chernyshev}},\ }\href
	{https://doi.org/10.1103/PhysRevLett.119.157201} {\bibfield  {journal}
		{\bibinfo  {journal} {Phys. Rev. Lett.}\ }\textbf {\bibinfo {volume} {119}},\
		\bibinfo {pages} {157201} (\bibinfo {year} {2017})}\BibitemShut {NoStop}%
	\bibitem [{\citenamefont {Zhang}\ \emph {et~al.}(2018)\citenamefont {Zhang},
		\citenamefont {Mahmood}, \citenamefont {Daum}, \citenamefont {Dun},
		\citenamefont {Paddison}, \citenamefont {Laurita}, \citenamefont {Hong},
		\citenamefont {Zhou}, \citenamefont {Armitage},\ and\ \citenamefont
		{Mourigal}}]{Zhang031001}%
	\BibitemOpen
	\bibfield  {author} {\bibinfo {author} {\bibfnamefont {X.}~\bibnamefont
			{Zhang}}, \bibinfo {author} {\bibfnamefont {F.}~\bibnamefont {Mahmood}},
		\bibinfo {author} {\bibfnamefont {M.}~\bibnamefont {Daum}}, \bibinfo {author}
		{\bibfnamefont {Z.}~\bibnamefont {Dun}}, \bibinfo {author} {\bibfnamefont
			{J.~A.~M.}\ \bibnamefont {Paddison}}, \bibinfo {author} {\bibfnamefont
			{N.~J.}\ \bibnamefont {Laurita}}, \bibinfo {author} {\bibfnamefont
			{T.}~\bibnamefont {Hong}}, \bibinfo {author} {\bibfnamefont {H.}~\bibnamefont
			{Zhou}}, \bibinfo {author} {\bibfnamefont {N.~P.}\ \bibnamefont {Armitage}},\
		and\ \bibinfo {author} {\bibfnamefont {M.}~\bibnamefont {Mourigal}},\ }\href
	{https://doi.org/10.1103/PhysRevX.8.031001} {\bibfield  {journal} {\bibinfo
			{journal} {Phys. Rev. X}\ }\textbf {\bibinfo {volume} {8}},\ \bibinfo {pages}
		{031001} (\bibinfo {year} {2018})}\BibitemShut {NoStop}%
	\bibitem [{\citenamefont {Ma}\ \emph {et~al.}(2018)\citenamefont {Ma},
		\citenamefont {Wang}, \citenamefont {Dong}, \citenamefont {Zhang},
		\citenamefont {Li}, \citenamefont {Zheng}, \citenamefont {Yu}, \citenamefont
		{Wang}, \citenamefont {Che}, \citenamefont {Ran}, \citenamefont {Bao},
		\citenamefont {Cai}, \citenamefont {\ifmmode~\check{C}\else
			\v{C}\fi{}erm\'ak}, \citenamefont {Schneidewind}, \citenamefont {Yano},
		\citenamefont {Gardner}, \citenamefont {Lu}, \citenamefont {Yu},
		\citenamefont {Liu}, \citenamefont {Li}, \citenamefont {Li},\ and\
		\citenamefont {Wen}}]{Ma087201}%
	\BibitemOpen
	\bibfield  {author} {\bibinfo {author} {\bibfnamefont {Z.}~\bibnamefont
			{Ma}}, \bibinfo {author} {\bibfnamefont {J.}~\bibnamefont {Wang}}, \bibinfo
		{author} {\bibfnamefont {Z.-Y.}\ \bibnamefont {Dong}}, \bibinfo {author}
		{\bibfnamefont {J.}~\bibnamefont {Zhang}}, \bibinfo {author} {\bibfnamefont
			{S.}~\bibnamefont {Li}}, \bibinfo {author} {\bibfnamefont {S.-H.}\
			\bibnamefont {Zheng}}, \bibinfo {author} {\bibfnamefont {Y.}~\bibnamefont
			{Yu}}, \bibinfo {author} {\bibfnamefont {W.}~\bibnamefont {Wang}}, \bibinfo
		{author} {\bibfnamefont {L.}~\bibnamefont {Che}}, \bibinfo {author}
		{\bibfnamefont {K.}~\bibnamefont {Ran}}, \bibinfo {author} {\bibfnamefont
			{S.}~\bibnamefont {Bao}}, \bibinfo {author} {\bibfnamefont {Z.}~\bibnamefont
			{Cai}}, \bibinfo {author} {\bibfnamefont {P.}~\bibnamefont
			{\ifmmode~\check{C}\else \v{C}\fi{}erm\'ak}}, \bibinfo {author}
		{\bibfnamefont {A.}~\bibnamefont {Schneidewind}}, \bibinfo {author}
		{\bibfnamefont {S.}~\bibnamefont {Yano}}, \bibinfo {author} {\bibfnamefont
			{J.~S.}\ \bibnamefont {Gardner}}, \bibinfo {author} {\bibfnamefont
			{X.}~\bibnamefont {Lu}}, \bibinfo {author} {\bibfnamefont {S.-L.}\
			\bibnamefont {Yu}}, \bibinfo {author} {\bibfnamefont {J.-M.}\ \bibnamefont
			{Liu}}, \bibinfo {author} {\bibfnamefont {S.}~\bibnamefont {Li}}, \bibinfo
		{author} {\bibfnamefont {J.-X.}\ \bibnamefont {Li}},\ and\ \bibinfo {author}
		{\bibfnamefont {J.}~\bibnamefont {Wen}},\ }\href
	{https://doi.org/10.1103/PhysRevLett.120.087201} {\bibfield  {journal}
		{\bibinfo  {journal} {Phys. Rev. Lett.}\ }\textbf {\bibinfo {volume} {120}},\
		\bibinfo {pages} {087201} (\bibinfo {year} {2018})}\BibitemShut {NoStop}%
	\bibitem [{\citenamefont {Liu}\ \emph {et~al.}(2018)\citenamefont {Liu},
		\citenamefont {Shao}, \citenamefont {Lin}, \citenamefont {Guo},\ and\
		\citenamefont {Sandvik}}]{Liu041040}%
	\BibitemOpen
	\bibfield  {author} {\bibinfo {author} {\bibfnamefont {L.}~\bibnamefont
			{Liu}}, \bibinfo {author} {\bibfnamefont {H.}~\bibnamefont {Shao}}, \bibinfo
		{author} {\bibfnamefont {Y.-C.}\ \bibnamefont {Lin}}, \bibinfo {author}
		{\bibfnamefont {W.}~\bibnamefont {Guo}},\ and\ \bibinfo {author}
		{\bibfnamefont {A.~W.}\ \bibnamefont {Sandvik}},\ }\href
	{https://doi.org/10.1103/PhysRevX.8.041040} {\bibfield  {journal} {\bibinfo
			{journal} {Phys. Rev. X}\ }\textbf {\bibinfo {volume} {8}},\ \bibinfo {pages}
		{041040} (\bibinfo {year} {2018})}\BibitemShut {NoStop}%
	\bibitem [{\citenamefont {Ma}\ \emph {et~al.}(2021)\citenamefont {Ma},
		\citenamefont {Dong}, \citenamefont {Wang}, \citenamefont {Zheng},
		\citenamefont {Ran}, \citenamefont {Bao}, \citenamefont {Cai}, \citenamefont
		{Shangguan}, \citenamefont {Wang}, \citenamefont {Boehm}, \citenamefont
		{Steffens}, \citenamefont {Regnault}, \citenamefont {Wang}, \citenamefont
		{Su}, \citenamefont {Yu}, \citenamefont {Liu}, \citenamefont {Li},\ and\
		\citenamefont {Wen}}]{Ma224433}%
	\BibitemOpen
	\bibfield  {author} {\bibinfo {author} {\bibfnamefont {Z.}~\bibnamefont
			{Ma}}, \bibinfo {author} {\bibfnamefont {Z.-Y.}\ \bibnamefont {Dong}},
		\bibinfo {author} {\bibfnamefont {J.}~\bibnamefont {Wang}}, \bibinfo {author}
		{\bibfnamefont {S.}~\bibnamefont {Zheng}}, \bibinfo {author} {\bibfnamefont
			{K.}~\bibnamefont {Ran}}, \bibinfo {author} {\bibfnamefont {S.}~\bibnamefont
			{Bao}}, \bibinfo {author} {\bibfnamefont {Z.}~\bibnamefont {Cai}}, \bibinfo
		{author} {\bibfnamefont {Y.}~\bibnamefont {Shangguan}}, \bibinfo {author}
		{\bibfnamefont {W.}~\bibnamefont {Wang}}, \bibinfo {author} {\bibfnamefont
			{M.}~\bibnamefont {Boehm}}, \bibinfo {author} {\bibfnamefont
			{P.}~\bibnamefont {Steffens}}, \bibinfo {author} {\bibfnamefont {L.-P.}\
			\bibnamefont {Regnault}}, \bibinfo {author} {\bibfnamefont {X.}~\bibnamefont
			{Wang}}, \bibinfo {author} {\bibfnamefont {Y.}~\bibnamefont {Su}}, \bibinfo
		{author} {\bibfnamefont {S.-L.}\ \bibnamefont {Yu}}, \bibinfo {author}
		{\bibfnamefont {J.-M.}\ \bibnamefont {Liu}}, \bibinfo {author} {\bibfnamefont
			{J.-X.}\ \bibnamefont {Li}},\ and\ \bibinfo {author} {\bibfnamefont
			{J.}~\bibnamefont {Wen}},\ }\href
	{https://doi.org/10.1103/PhysRevB.104.224433} {\bibfield  {journal} {\bibinfo
			{journal} {Phys. Rev. B}\ }\textbf {\bibinfo {volume} {104}},\ \bibinfo
		{pages} {224433} (\bibinfo {year} {2021})}\BibitemShut {NoStop}%
	\bibitem [{\citenamefont {Paddison}\ \emph {et~al.}(2017)\citenamefont
		{Paddison}, \citenamefont {Daum}, \citenamefont {Dun}, \citenamefont
		{Ehlers}, \citenamefont {Liu}, \citenamefont {Stone}, \citenamefont {Zhou},\
		and\ \citenamefont {Mourigal}}]{Paddison117}%
	\BibitemOpen
	\bibfield  {author} {\bibinfo {author} {\bibfnamefont {J.~A.~M.}\
			\bibnamefont {Paddison}}, \bibinfo {author} {\bibfnamefont {M.}~\bibnamefont
			{Daum}}, \bibinfo {author} {\bibfnamefont {Z.}~\bibnamefont {Dun}}, \bibinfo
		{author} {\bibfnamefont {G.}~\bibnamefont {Ehlers}}, \bibinfo {author}
		{\bibfnamefont {Y.}~\bibnamefont {Liu}}, \bibinfo {author} {\bibfnamefont
			{M.~B.}\ \bibnamefont {Stone}}, \bibinfo {author} {\bibfnamefont
			{H.}~\bibnamefont {Zhou}},\ and\ \bibinfo {author} {\bibfnamefont
			{M.}~\bibnamefont {Mourigal}},\ }\href {https://doi.org/10.1038/nphys3971}
	{\bibfield  {journal} {\bibinfo  {journal} {Nat. Phys}\ }\textbf {\bibinfo
			{volume} {13}},\ \bibinfo {pages} {117} (\bibinfo {year} {2017})}\BibitemShut
	{NoStop}%
	\bibitem [{\citenamefont {Li}\ \emph {et~al.}(2016{\natexlab{b}})\citenamefont
		{Li}, \citenamefont {Adroja}, \citenamefont {Biswas}, \citenamefont {Baker},
		\citenamefont {Zhang}, \citenamefont {Liu}, \citenamefont {Tsirlin},
		\citenamefont {Gegenwart},\ and\ \citenamefont {Zhang}}]{Li097201}%
	\BibitemOpen
	\bibfield  {author} {\bibinfo {author} {\bibfnamefont {Y.}~\bibnamefont
			{Li}}, \bibinfo {author} {\bibfnamefont {D.}~\bibnamefont {Adroja}}, \bibinfo
		{author} {\bibfnamefont {P.~K.}\ \bibnamefont {Biswas}}, \bibinfo {author}
		{\bibfnamefont {P.~J.}\ \bibnamefont {Baker}}, \bibinfo {author}
		{\bibfnamefont {Q.}~\bibnamefont {Zhang}}, \bibinfo {author} {\bibfnamefont
			{J.}~\bibnamefont {Liu}}, \bibinfo {author} {\bibfnamefont {A.~A.}\
			\bibnamefont {Tsirlin}}, \bibinfo {author} {\bibfnamefont {P.}~\bibnamefont
			{Gegenwart}},\ and\ \bibinfo {author} {\bibfnamefont {Q.}~\bibnamefont
			{Zhang}},\ }\href {https://doi.org/10.1103/PhysRevLett.117.097201} {\bibfield
		{journal} {\bibinfo  {journal} {Phys. Rev. Lett.}\ }\textbf {\bibinfo
			{volume} {117}},\ \bibinfo {pages} {097201} (\bibinfo {year}
		{2016}{\natexlab{b}})}\BibitemShut {NoStop}%
	\bibitem [{\citenamefont {Schmidt}\ \emph {et~al.}(2021)\citenamefont
		{Schmidt}, \citenamefont {Sichelschmidt}, \citenamefont {Ranjith},
		\citenamefont {Doert},\ and\ \citenamefont {Baenitz}}]{Bchmidt214445}%
	\BibitemOpen
	\bibfield  {author} {\bibinfo {author} {\bibfnamefont {B.}~\bibnamefont
			{Schmidt}}, \bibinfo {author} {\bibfnamefont {J.}~\bibnamefont
			{Sichelschmidt}}, \bibinfo {author} {\bibfnamefont {K.~M.}\ \bibnamefont
			{Ranjith}}, \bibinfo {author} {\bibfnamefont {T.}~\bibnamefont {Doert}},\
		and\ \bibinfo {author} {\bibfnamefont {M.}~\bibnamefont {Baenitz}},\ }\href
	{https://doi.org/10.1103/PhysRevB.103.214445} {\bibfield  {journal} {\bibinfo
			{journal} {Phys. Rev. B}\ }\textbf {\bibinfo {volume} {103}},\ \bibinfo
		{pages} {214445} (\bibinfo {year} {2021})}\BibitemShut {NoStop}%
	\bibitem [{\citenamefont {Baenitz}\ \emph {et~al.}(2018)\citenamefont
		{Baenitz}, \citenamefont {Schlender}, \citenamefont {Sichelschmidt},
		\citenamefont {Onykiienko}, \citenamefont {Zangeneh}, \citenamefont
		{Ranjith}, \citenamefont {Sarkar}, \citenamefont {Hozoi}, \citenamefont
		{Walker}, \citenamefont {Orain}, \citenamefont {Yasuoka}, \citenamefont
		{van~den Brink}, \citenamefont {Klauss}, \citenamefont {Inosov},\ and\
		\citenamefont {Doert}}]{Baenitz220409}%
	\BibitemOpen
	\bibfield  {author} {\bibinfo {author} {\bibfnamefont {M.}~\bibnamefont
			{Baenitz}}, \bibinfo {author} {\bibfnamefont {P.}~\bibnamefont {Schlender}},
		\bibinfo {author} {\bibfnamefont {J.}~\bibnamefont {Sichelschmidt}}, \bibinfo
		{author} {\bibfnamefont {Y.~A.}\ \bibnamefont {Onykiienko}}, \bibinfo
		{author} {\bibfnamefont {Z.}~\bibnamefont {Zangeneh}}, \bibinfo {author}
		{\bibfnamefont {K.~M.}\ \bibnamefont {Ranjith}}, \bibinfo {author}
		{\bibfnamefont {R.}~\bibnamefont {Sarkar}}, \bibinfo {author} {\bibfnamefont
			{L.}~\bibnamefont {Hozoi}}, \bibinfo {author} {\bibfnamefont {H.~C.}\
			\bibnamefont {Walker}}, \bibinfo {author} {\bibfnamefont {J.-C.}\
			\bibnamefont {Orain}}, \bibinfo {author} {\bibfnamefont {H.}~\bibnamefont
			{Yasuoka}}, \bibinfo {author} {\bibfnamefont {J.}~\bibnamefont {van~den
				Brink}}, \bibinfo {author} {\bibfnamefont {H.~H.}\ \bibnamefont {Klauss}},
		\bibinfo {author} {\bibfnamefont {D.~S.}\ \bibnamefont {Inosov}},\ and\
		\bibinfo {author} {\bibfnamefont {T.}~\bibnamefont {Doert}},\ }\href
	{https://doi.org/10.1103/PhysRevB.98.220409} {\bibfield  {journal} {\bibinfo
			{journal} {Phys. Rev. B}\ }\textbf {\bibinfo {volume} {98}},\ \bibinfo
		{pages} {220409} (\bibinfo {year} {2018})}\BibitemShut {NoStop}%
	\bibitem [{\citenamefont {Ranjith}\ \emph
		{et~al.}(2019{\natexlab{a}})\citenamefont {Ranjith}, \citenamefont
		{Dmytriieva}, \citenamefont {Khim}, \citenamefont {Sichelschmidt},
		\citenamefont {Luther}, \citenamefont {Ehlers}, \citenamefont {Yasuoka},
		\citenamefont {Wosnitza}, \citenamefont {Tsirlin}, \citenamefont {K\"uhne},\
		and\ \citenamefont {Baenitz}}]{Ranjith180401}%
	\BibitemOpen
	\bibfield  {author} {\bibinfo {author} {\bibfnamefont {K.~M.}\ \bibnamefont
			{Ranjith}}, \bibinfo {author} {\bibfnamefont {D.}~\bibnamefont {Dmytriieva}},
		\bibinfo {author} {\bibfnamefont {S.}~\bibnamefont {Khim}}, \bibinfo {author}
		{\bibfnamefont {J.}~\bibnamefont {Sichelschmidt}}, \bibinfo {author}
		{\bibfnamefont {S.}~\bibnamefont {Luther}}, \bibinfo {author} {\bibfnamefont
			{D.}~\bibnamefont {Ehlers}}, \bibinfo {author} {\bibfnamefont
			{H.}~\bibnamefont {Yasuoka}}, \bibinfo {author} {\bibfnamefont
			{J.}~\bibnamefont {Wosnitza}}, \bibinfo {author} {\bibfnamefont {A.~A.}\
			\bibnamefont {Tsirlin}}, \bibinfo {author} {\bibfnamefont {H.}~\bibnamefont
			{K\"uhne}},\ and\ \bibinfo {author} {\bibfnamefont {M.}~\bibnamefont
			{Baenitz}},\ }\href {https://doi.org/10.1103/PhysRevB.99.180401} {\bibfield
		{journal} {\bibinfo  {journal} {Phys. Rev. B}\ }\textbf {\bibinfo {volume}
			{99}},\ \bibinfo {pages} {180401} (\bibinfo {year}
		{2019}{\natexlab{a}})}\BibitemShut {NoStop}%
	\bibitem [{\citenamefont {Bordelon}\ \emph {et~al.}(2019)\citenamefont
		{Bordelon}, \citenamefont {Kenney}, \citenamefont {Liu}, \citenamefont
		{Hogan}, \citenamefont {Posthuma}, \citenamefont {Kavand}, \citenamefont
		{Lyu}, \citenamefont {Sherwin}, \citenamefont {Butch}, \citenamefont {Brown},
		\citenamefont {Graf}, \citenamefont {Balents},\ and\ \citenamefont
		{Wilson}}]{Bordelon1058}%
	\BibitemOpen
	\bibfield  {author} {\bibinfo {author} {\bibfnamefont {M.~M.}\ \bibnamefont
			{Bordelon}}, \bibinfo {author} {\bibfnamefont {E.}~\bibnamefont {Kenney}},
		\bibinfo {author} {\bibfnamefont {C.}~\bibnamefont {Liu}}, \bibinfo {author}
		{\bibfnamefont {T.}~\bibnamefont {Hogan}}, \bibinfo {author} {\bibfnamefont
			{L.}~\bibnamefont {Posthuma}}, \bibinfo {author} {\bibfnamefont
			{M.}~\bibnamefont {Kavand}}, \bibinfo {author} {\bibfnamefont
			{Y.}~\bibnamefont {Lyu}}, \bibinfo {author} {\bibfnamefont {M.}~\bibnamefont
			{Sherwin}}, \bibinfo {author} {\bibfnamefont {N.~P.}\ \bibnamefont {Butch}},
		\bibinfo {author} {\bibfnamefont {C.}~\bibnamefont {Brown}}, \bibinfo
		{author} {\bibfnamefont {M.~J.}\ \bibnamefont {Graf}}, \bibinfo {author}
		{\bibfnamefont {L.}~\bibnamefont {Balents}},\ and\ \bibinfo {author}
		{\bibfnamefont {S.~D.}\ \bibnamefont {Wilson}},\ }\href
	{https://doi.org/10.1038/s41567-019-0594-5} {\bibfield  {journal} {\bibinfo
			{journal} {Nat. Phys}\ }\textbf {\bibinfo {volume} {15}},\ \bibinfo {pages}
		{1058} (\bibinfo {year} {2019})}\BibitemShut {NoStop}%
	\bibitem [{\citenamefont {Ding}\ \emph {et~al.}(2019)\citenamefont {Ding},
		\citenamefont {Manuel}, \citenamefont {Bachus}, \citenamefont {Gru\ss{}ler},
		\citenamefont {Gegenwart}, \citenamefont {Singleton}, \citenamefont
		{Johnson}, \citenamefont {Walker}, \citenamefont {Adroja}, \citenamefont
		{Hillier},\ and\ \citenamefont {Tsirlin}}]{Ding144432}%
	\BibitemOpen
	\bibfield  {author} {\bibinfo {author} {\bibfnamefont {L.}~\bibnamefont
			{Ding}}, \bibinfo {author} {\bibfnamefont {P.}~\bibnamefont {Manuel}},
		\bibinfo {author} {\bibfnamefont {S.}~\bibnamefont {Bachus}}, \bibinfo
		{author} {\bibfnamefont {F.}~\bibnamefont {Gru\ss{}ler}}, \bibinfo {author}
		{\bibfnamefont {P.}~\bibnamefont {Gegenwart}}, \bibinfo {author}
		{\bibfnamefont {J.}~\bibnamefont {Singleton}}, \bibinfo {author}
		{\bibfnamefont {R.~D.}\ \bibnamefont {Johnson}}, \bibinfo {author}
		{\bibfnamefont {H.~C.}\ \bibnamefont {Walker}}, \bibinfo {author}
		{\bibfnamefont {D.~T.}\ \bibnamefont {Adroja}}, \bibinfo {author}
		{\bibfnamefont {A.~D.}\ \bibnamefont {Hillier}},\ and\ \bibinfo {author}
		{\bibfnamefont {A.~A.}\ \bibnamefont {Tsirlin}},\ }\href
	{https://doi.org/10.1103/PhysRevB.100.144432} {\bibfield  {journal} {\bibinfo
			{journal} {Phys. Rev. B}\ }\textbf {\bibinfo {volume} {100}},\ \bibinfo
		{pages} {144432} (\bibinfo {year} {2019})}\BibitemShut {NoStop}%
	\bibitem [{\citenamefont {Ranjith}\ \emph
		{et~al.}(2019{\natexlab{b}})\citenamefont {Ranjith}, \citenamefont {Luther},
		\citenamefont {Reimann}, \citenamefont {Schmidt}, \citenamefont {Schlender},
		\citenamefont {Sichelschmidt}, \citenamefont {Yasuoka}, \citenamefont
		{Strydom}, \citenamefont {Skourski}, \citenamefont {Wosnitza}, \citenamefont
		{K\"uhne}, \citenamefont {Doert},\ and\ \citenamefont
		{Baenitz}}]{Ranjith224417}%
	\BibitemOpen
	\bibfield  {author} {\bibinfo {author} {\bibfnamefont {K.~M.}\ \bibnamefont
			{Ranjith}}, \bibinfo {author} {\bibfnamefont {S.}~\bibnamefont {Luther}},
		\bibinfo {author} {\bibfnamefont {T.}~\bibnamefont {Reimann}}, \bibinfo
		{author} {\bibfnamefont {B.}~\bibnamefont {Schmidt}}, \bibinfo {author}
		{\bibfnamefont {P.}~\bibnamefont {Schlender}}, \bibinfo {author}
		{\bibfnamefont {J.}~\bibnamefont {Sichelschmidt}}, \bibinfo {author}
		{\bibfnamefont {H.}~\bibnamefont {Yasuoka}}, \bibinfo {author} {\bibfnamefont
			{A.~M.}\ \bibnamefont {Strydom}}, \bibinfo {author} {\bibfnamefont
			{Y.}~\bibnamefont {Skourski}}, \bibinfo {author} {\bibfnamefont
			{J.}~\bibnamefont {Wosnitza}}, \bibinfo {author} {\bibfnamefont
			{H.}~\bibnamefont {K\"uhne}}, \bibinfo {author} {\bibfnamefont
			{T.}~\bibnamefont {Doert}},\ and\ \bibinfo {author} {\bibfnamefont
			{M.}~\bibnamefont {Baenitz}},\ }\href
	{https://doi.org/10.1103/PhysRevB.100.224417} {\bibfield  {journal} {\bibinfo
			{journal} {Phys. Rev. B}\ }\textbf {\bibinfo {volume} {100}},\ \bibinfo
		{pages} {224417} (\bibinfo {year} {2019}{\natexlab{b}})}\BibitemShut
	{NoStop}%
	\bibitem [{\citenamefont {Guo}\ \emph {et~al.}(2020{\natexlab{a}})\citenamefont
		{Guo}, \citenamefont {Zhao}, \citenamefont {Ohira-Kawamura}, \citenamefont
		{Ling}, \citenamefont {Wang}, \citenamefont {He}, \citenamefont {Nakajima},
		\citenamefont {Li},\ and\ \citenamefont {Zhang}}]{Guo064410}%
	\BibitemOpen
	\bibfield  {author} {\bibinfo {author} {\bibfnamefont {J.}~\bibnamefont
			{Guo}}, \bibinfo {author} {\bibfnamefont {X.}~\bibnamefont {Zhao}}, \bibinfo
		{author} {\bibfnamefont {S.}~\bibnamefont {Ohira-Kawamura}}, \bibinfo
		{author} {\bibfnamefont {L.}~\bibnamefont {Ling}}, \bibinfo {author}
		{\bibfnamefont {J.}~\bibnamefont {Wang}}, \bibinfo {author} {\bibfnamefont
			{L.}~\bibnamefont {He}}, \bibinfo {author} {\bibfnamefont {K.}~\bibnamefont
			{Nakajima}}, \bibinfo {author} {\bibfnamefont {B.}~\bibnamefont {Li}},\ and\
		\bibinfo {author} {\bibfnamefont {Z.}~\bibnamefont {Zhang}},\ }\href
	{https://doi.org/10.1103/PhysRevMaterials.4.064410} {\bibfield  {journal}
		{\bibinfo  {journal} {Phys. Rev. Mater.}\ }\textbf {\bibinfo {volume} {4}},\
		\bibinfo {pages} {064410} (\bibinfo {year} {2020}{\natexlab{a}})}\BibitemShut
	{NoStop}%
	\bibitem [{\citenamefont {Bordelon}\ \emph {et~al.}(2020)\citenamefont
		{Bordelon}, \citenamefont {Liu}, \citenamefont {Posthuma}, \citenamefont
		{Sarte}, \citenamefont {Butch}, \citenamefont {Pajerowski}, \citenamefont
		{Banerjee}, \citenamefont {Balents},\ and\ \citenamefont
		{Wilson}}]{Bordelon224427}%
	\BibitemOpen
	\bibfield  {author} {\bibinfo {author} {\bibfnamefont {M.~M.}\ \bibnamefont
			{Bordelon}}, \bibinfo {author} {\bibfnamefont {C.}~\bibnamefont {Liu}},
		\bibinfo {author} {\bibfnamefont {L.}~\bibnamefont {Posthuma}}, \bibinfo
		{author} {\bibfnamefont {P.~M.}\ \bibnamefont {Sarte}}, \bibinfo {author}
		{\bibfnamefont {N.~P.}\ \bibnamefont {Butch}}, \bibinfo {author}
		{\bibfnamefont {D.~M.}\ \bibnamefont {Pajerowski}}, \bibinfo {author}
		{\bibfnamefont {A.}~\bibnamefont {Banerjee}}, \bibinfo {author}
		{\bibfnamefont {L.}~\bibnamefont {Balents}},\ and\ \bibinfo {author}
		{\bibfnamefont {S.~D.}\ \bibnamefont {Wilson}},\ }\href
	{https://doi.org/10.1103/PhysRevB.101.224427} {\bibfield  {journal} {\bibinfo
			{journal} {Phys. Rev. B}\ }\textbf {\bibinfo {volume} {101}},\ \bibinfo
		{pages} {224427} (\bibinfo {year} {2020})}\BibitemShut {NoStop}%
	\bibitem [{\citenamefont {Zhang}\ \emph {et~al.}(2022)\citenamefont {Zhang},
		\citenamefont {Li}, \citenamefont {Xie}, \citenamefont {Zhuo}, \citenamefont
		{Adroja}, \citenamefont {Baker}, \citenamefont {Perring}, \citenamefont
		{Zhang}, \citenamefont {Jin}, \citenamefont {Ji}, \citenamefont {Wang},
		\citenamefont {Ma},\ and\ \citenamefont {Zhang}}]{Zhang085115}%
	\BibitemOpen
	\bibfield  {author} {\bibinfo {author} {\bibfnamefont {Z.}~\bibnamefont
			{Zhang}}, \bibinfo {author} {\bibfnamefont {J.}~\bibnamefont {Li}}, \bibinfo
		{author} {\bibfnamefont {M.}~\bibnamefont {Xie}}, \bibinfo {author}
		{\bibfnamefont {W.}~\bibnamefont {Zhuo}}, \bibinfo {author} {\bibfnamefont
			{D.~T.}\ \bibnamefont {Adroja}}, \bibinfo {author} {\bibfnamefont {P.~J.}\
			\bibnamefont {Baker}}, \bibinfo {author} {\bibfnamefont {T.~G.}\ \bibnamefont
			{Perring}}, \bibinfo {author} {\bibfnamefont {A.}~\bibnamefont {Zhang}},
		\bibinfo {author} {\bibfnamefont {F.}~\bibnamefont {Jin}}, \bibinfo {author}
		{\bibfnamefont {J.}~\bibnamefont {Ji}}, \bibinfo {author} {\bibfnamefont
			{X.}~\bibnamefont {Wang}}, \bibinfo {author} {\bibfnamefont {J.}~\bibnamefont
			{Ma}},\ and\ \bibinfo {author} {\bibfnamefont {Q.}~\bibnamefont {Zhang}},\
	}\href {https://doi.org/10.1103/PhysRevB.106.085115} {\bibfield  {journal}
		{\bibinfo  {journal} {Phys. Rev. B}\ }\textbf {\bibinfo {volume} {106}},\
		\bibinfo {pages} {085115} (\bibinfo {year} {2022})}\BibitemShut {NoStop}%
	\bibitem [{\citenamefont {Dai}\ \emph {et~al.}(2021)\citenamefont {Dai},
		\citenamefont {Zhang}, \citenamefont {Xie}, \citenamefont {Duan},
		\citenamefont {Gao}, \citenamefont {Zhu}, \citenamefont {Feng}, \citenamefont
		{Tao}, \citenamefont {Huang}, \citenamefont {Cao}, \citenamefont
		{Podlesnyak}, \citenamefont {Granroth}, \citenamefont {Everett},
		\citenamefont {Neuefeind}, \citenamefont {Voneshen}, \citenamefont {Wang},
		\citenamefont {Tan}, \citenamefont {Morosan}, \citenamefont {Wang},
		\citenamefont {Lin}, \citenamefont {Shu}, \citenamefont {Chen}, \citenamefont
		{Guo}, \citenamefont {Lu},\ and\ \citenamefont {Dai}}]{Dai021044}%
	\BibitemOpen
	\bibfield  {author} {\bibinfo {author} {\bibfnamefont {P.-L.}\ \bibnamefont
			{Dai}}, \bibinfo {author} {\bibfnamefont {G.}~\bibnamefont {Zhang}}, \bibinfo
		{author} {\bibfnamefont {Y.}~\bibnamefont {Xie}}, \bibinfo {author}
		{\bibfnamefont {C.}~\bibnamefont {Duan}}, \bibinfo {author} {\bibfnamefont
			{Y.}~\bibnamefont {Gao}}, \bibinfo {author} {\bibfnamefont {Z.}~\bibnamefont
			{Zhu}}, \bibinfo {author} {\bibfnamefont {E.}~\bibnamefont {Feng}}, \bibinfo
		{author} {\bibfnamefont {Z.}~\bibnamefont {Tao}}, \bibinfo {author}
		{\bibfnamefont {C.-L.}\ \bibnamefont {Huang}}, \bibinfo {author}
		{\bibfnamefont {H.}~\bibnamefont {Cao}}, \bibinfo {author} {\bibfnamefont
			{A.}~\bibnamefont {Podlesnyak}}, \bibinfo {author} {\bibfnamefont {G.~E.}\
			\bibnamefont {Granroth}}, \bibinfo {author} {\bibfnamefont {M.~S.}\
			\bibnamefont {Everett}}, \bibinfo {author} {\bibfnamefont {J.~C.}\
			\bibnamefont {Neuefeind}}, \bibinfo {author} {\bibfnamefont {D.}~\bibnamefont
			{Voneshen}}, \bibinfo {author} {\bibfnamefont {S.}~\bibnamefont {Wang}},
		\bibinfo {author} {\bibfnamefont {G.}~\bibnamefont {Tan}}, \bibinfo {author}
		{\bibfnamefont {E.}~\bibnamefont {Morosan}}, \bibinfo {author} {\bibfnamefont
			{X.}~\bibnamefont {Wang}}, \bibinfo {author} {\bibfnamefont {H.-Q.}\
			\bibnamefont {Lin}}, \bibinfo {author} {\bibfnamefont {L.}~\bibnamefont
			{Shu}}, \bibinfo {author} {\bibfnamefont {G.}~\bibnamefont {Chen}}, \bibinfo
		{author} {\bibfnamefont {Y.}~\bibnamefont {Guo}}, \bibinfo {author}
		{\bibfnamefont {X.}~\bibnamefont {Lu}},\ and\ \bibinfo {author}
		{\bibfnamefont {P.}~\bibnamefont {Dai}},\ }\href
	{https://doi.org/10.1103/PhysRevX.11.021044} {\bibfield  {journal} {\bibinfo
			{journal} {Phys. Rev. X}\ }\textbf {\bibinfo {volume} {11}},\ \bibinfo
		{pages} {021044} (\bibinfo {year} {2021})}\BibitemShut {NoStop}%
	\bibitem [{\citenamefont {Sarkar}\ \emph {et~al.}(2019)\citenamefont {Sarkar},
		\citenamefont {Schlender}, \citenamefont {Grinenko}, \citenamefont
		{Haeussler}, \citenamefont {Baker}, \citenamefont {Doert},\ and\
		\citenamefont {Klauss}}]{Sarkar241116}%
	\BibitemOpen
	\bibfield  {author} {\bibinfo {author} {\bibfnamefont {R.}~\bibnamefont
			{Sarkar}}, \bibinfo {author} {\bibfnamefont {P.}~\bibnamefont {Schlender}},
		\bibinfo {author} {\bibfnamefont {V.}~\bibnamefont {Grinenko}}, \bibinfo
		{author} {\bibfnamefont {E.}~\bibnamefont {Haeussler}}, \bibinfo {author}
		{\bibfnamefont {P.~J.}\ \bibnamefont {Baker}}, \bibinfo {author}
		{\bibfnamefont {T.}~\bibnamefont {Doert}},\ and\ \bibinfo {author}
		{\bibfnamefont {H.-H.}\ \bibnamefont {Klauss}},\ }\href
	{https://doi.org/10.1103/PhysRevB.100.241116} {\bibfield  {journal} {\bibinfo
			{journal} {Phys. Rev. B}\ }\textbf {\bibinfo {volume} {100}},\ \bibinfo
		{pages} {241116} (\bibinfo {year} {2019})}\BibitemShut {NoStop}%
	\bibitem [{\citenamefont {Zeng}\ \emph {et~al.}(2020)\citenamefont {Zeng},
		\citenamefont {Ma}, \citenamefont {Gao}, \citenamefont {Tian}, \citenamefont
		{Ling},\ and\ \citenamefont {Pi}}]{Zeng045149}%
	\BibitemOpen
	\bibfield  {author} {\bibinfo {author} {\bibfnamefont {K.~Y.}\ \bibnamefont
			{Zeng}}, \bibinfo {author} {\bibfnamefont {L.}~\bibnamefont {Ma}}, \bibinfo
		{author} {\bibfnamefont {Y.~X.}\ \bibnamefont {Gao}}, \bibinfo {author}
		{\bibfnamefont {Z.~M.}\ \bibnamefont {Tian}}, \bibinfo {author}
		{\bibfnamefont {L.~S.}\ \bibnamefont {Ling}},\ and\ \bibinfo {author}
		{\bibfnamefont {L.}~\bibnamefont {Pi}},\ }\href
	{https://doi.org/10.1103/PhysRevB.102.045149} {\bibfield  {journal} {\bibinfo
			{journal} {Phys. Rev. B}\ }\textbf {\bibinfo {volume} {102}},\ \bibinfo
		{pages} {045149} (\bibinfo {year} {2020})}\BibitemShut {NoStop}%
	\bibitem [{\citenamefont {Bag}\ \emph {et~al.}(2021)\citenamefont {Bag},
		\citenamefont {Ennis}, \citenamefont {Liu}, \citenamefont {Dissanayake},
		\citenamefont {Shi}, \citenamefont {Liu}, \citenamefont {Balents},\ and\
		\citenamefont {Haravifard}}]{BagL220403}%
	\BibitemOpen
	\bibfield  {author} {\bibinfo {author} {\bibfnamefont {R.}~\bibnamefont
			{Bag}}, \bibinfo {author} {\bibfnamefont {M.}~\bibnamefont {Ennis}}, \bibinfo
		{author} {\bibfnamefont {C.}~\bibnamefont {Liu}}, \bibinfo {author}
		{\bibfnamefont {S.~E.}\ \bibnamefont {Dissanayake}}, \bibinfo {author}
		{\bibfnamefont {Z.}~\bibnamefont {Shi}}, \bibinfo {author} {\bibfnamefont
			{J.}~\bibnamefont {Liu}}, \bibinfo {author} {\bibfnamefont {L.}~\bibnamefont
			{Balents}},\ and\ \bibinfo {author} {\bibfnamefont {S.}~\bibnamefont
			{Haravifard}},\ }\href {https://doi.org/10.1103/PhysRevB.104.L220403}
	{\bibfield  {journal} {\bibinfo  {journal} {Phys. Rev. B}\ }\textbf {\bibinfo
			{volume} {104}},\ \bibinfo {pages} {L220403} (\bibinfo {year}
		{2021})}\BibitemShut {NoStop}%
	\bibitem [{\citenamefont {Chadeyron}\ \emph {et~al.}(1997)\citenamefont
		{Chadeyron}, \citenamefont {El-Ghozzi}, \citenamefont {Mahiou}, \citenamefont
		{Arbus},\ and\ \citenamefont {Cousseins}}]{Chadeyron261}%
	\BibitemOpen
	\bibfield  {author} {\bibinfo {author} {\bibfnamefont {G.}~\bibnamefont
			{Chadeyron}}, \bibinfo {author} {\bibfnamefont {M.}~\bibnamefont
			{El-Ghozzi}}, \bibinfo {author} {\bibfnamefont {R.}~\bibnamefont {Mahiou}},
		\bibinfo {author} {\bibfnamefont {A.}~\bibnamefont {Arbus}},\ and\ \bibinfo
		{author} {\bibfnamefont {J.}~\bibnamefont {Cousseins}},\ }\href
	{https://doi.org/https://doi.org/10.1006/jssc.1996.7207} {\bibfield
		{journal} {\bibinfo  {journal} {J. Solid State Chem.}\ }\textbf {\bibinfo
			{volume} {128}},\ \bibinfo {pages} {261} (\bibinfo {year}
		{1997})}\BibitemShut {NoStop}%
	\bibitem [{sup()}]{supplementary}%
	\BibitemOpen
	\href@noop {} {}\bibinfo {note} {See Supplemental Material at http: LINK for
		additional information, which includes
		Refs.~\cite{Rauch174404,Abragam2012,Ranjith180401,Amato093301,Chadeyron261,Somesh104422,Orbach458}}\BibitemShut
	{NoStop}%
	\bibitem [{\citenamefont {Ranjith}\ \emph {et~al.}(2017)\citenamefont
		{Ranjith}, \citenamefont {Brinda}, \citenamefont {Arjun}, \citenamefont
		{Hegde},\ and\ \citenamefont {Nath}}]{Ranjith115804}%
	\BibitemOpen
	\bibfield  {author} {\bibinfo {author} {\bibfnamefont {K.~M.}\ \bibnamefont
			{Ranjith}}, \bibinfo {author} {\bibfnamefont {K.}~\bibnamefont {Brinda}},
		\bibinfo {author} {\bibfnamefont {U.}~\bibnamefont {Arjun}}, \bibinfo
		{author} {\bibfnamefont {N.~G.}\ \bibnamefont {Hegde}},\ and\ \bibinfo
		{author} {\bibfnamefont {R.}~\bibnamefont {Nath}},\ }\href
	{https://doi.org/10.1088/1361-648x/aa57be} {\bibfield  {journal} {\bibinfo
			{journal} {J. Phys.: Condens. Matter}\ }\textbf {\bibinfo {volume} {29}},\
		\bibinfo {pages} {115804} (\bibinfo {year} {2017})}\BibitemShut {NoStop}%
	\bibitem [{\citenamefont {Guo}\ \emph {et~al.}(2019)\citenamefont {Guo},
		\citenamefont {Ghasemi}, \citenamefont {Broholm},\ and\ \citenamefont
		{Cava}}]{Guo094404}%
	\BibitemOpen
	\bibfield  {author} {\bibinfo {author} {\bibfnamefont {S.}~\bibnamefont
			{Guo}}, \bibinfo {author} {\bibfnamefont {A.}~\bibnamefont {Ghasemi}},
		\bibinfo {author} {\bibfnamefont {C.~L.}\ \bibnamefont {Broholm}},\ and\
		\bibinfo {author} {\bibfnamefont {R.~J.}\ \bibnamefont {Cava}},\ }\href
	{https://doi.org/10.1103/PhysRevMaterials.3.094404} {\bibfield  {journal}
		{\bibinfo  {journal} {Phys. Rev. Mater.}\ }\textbf {\bibinfo {volume} {3}},\
		\bibinfo {pages} {094404} (\bibinfo {year} {2019})}\BibitemShut {NoStop}%
	\bibitem [{\citenamefont {Sibille}\ \emph {et~al.}(2015)\citenamefont
		{Sibille}, \citenamefont {Lhotel}, \citenamefont {Pomjakushin}, \citenamefont
		{Baines}, \citenamefont {Fennell},\ and\ \citenamefont
		{Kenzelmann}}]{Sibille097202}%
	\BibitemOpen
	\bibfield  {author} {\bibinfo {author} {\bibfnamefont {R.}~\bibnamefont
			{Sibille}}, \bibinfo {author} {\bibfnamefont {E.}~\bibnamefont {Lhotel}},
		\bibinfo {author} {\bibfnamefont {V.}~\bibnamefont {Pomjakushin}}, \bibinfo
		{author} {\bibfnamefont {C.}~\bibnamefont {Baines}}, \bibinfo {author}
		{\bibfnamefont {T.}~\bibnamefont {Fennell}},\ and\ \bibinfo {author}
		{\bibfnamefont {M.}~\bibnamefont {Kenzelmann}},\ }\href
	{https://doi.org/10.1103/PhysRevLett.115.097202} {\bibfield  {journal}
		{\bibinfo  {journal} {Phys. Rev. Lett.}\ }\textbf {\bibinfo {volume} {115}},\
		\bibinfo {pages} {097202} (\bibinfo {year} {2015})}\BibitemShut {NoStop}%
	\bibitem [{\citenamefont {Kittel}(2005)}]{Kittelc2005}%
	\BibitemOpen
	\bibfield  {author} {\bibinfo {author} {\bibfnamefont {C.}~\bibnamefont
			{Kittel}},\ }\href@noop {} {\emph {\bibinfo {title} {{Introduction to solid
					state physics}}}}\ (\bibinfo  {publisher} {J. Wiley},\ \bibinfo {address}
	{Hoboken, NJ},\ \bibinfo {year} {c2005})\ \bibinfo {note}
	{500317}\BibitemShut {NoStop}%
	\bibitem [{\citenamefont {Kundu}\ \emph {et~al.}(2020)\citenamefont {Kundu},
		\citenamefont {Hossain}, \citenamefont {S.}, \citenamefont {Das},
		\citenamefont {Baenitz}, \citenamefont {Baker}, \citenamefont {Orain},
		\citenamefont {Joshi}, \citenamefont {Mathieu}, \citenamefont {Mahadevan},
		\citenamefont {Pujari}, \citenamefont {Bhattacharjee}, \citenamefont
		{Mahajan},\ and\ \citenamefont {Sarma}}]{Kundu117206}%
	\BibitemOpen
	\bibfield  {author} {\bibinfo {author} {\bibfnamefont {S.}~\bibnamefont
			{Kundu}}, \bibinfo {author} {\bibfnamefont {A.}~\bibnamefont {Hossain}},
		\bibinfo {author} {\bibfnamefont {P.~K.}\ \bibnamefont {S.}}, \bibinfo
		{author} {\bibfnamefont {R.}~\bibnamefont {Das}}, \bibinfo {author}
		{\bibfnamefont {M.}~\bibnamefont {Baenitz}}, \bibinfo {author} {\bibfnamefont
			{P.~J.}\ \bibnamefont {Baker}}, \bibinfo {author} {\bibfnamefont {J.-C.}\
			\bibnamefont {Orain}}, \bibinfo {author} {\bibfnamefont {D.~C.}\ \bibnamefont
			{Joshi}}, \bibinfo {author} {\bibfnamefont {R.}~\bibnamefont {Mathieu}},
		\bibinfo {author} {\bibfnamefont {P.}~\bibnamefont {Mahadevan}}, \bibinfo
		{author} {\bibfnamefont {S.}~\bibnamefont {Pujari}}, \bibinfo {author}
		{\bibfnamefont {S.}~\bibnamefont {Bhattacharjee}}, \bibinfo {author}
		{\bibfnamefont {A.~V.}\ \bibnamefont {Mahajan}},\ and\ \bibinfo {author}
		{\bibfnamefont {D.~D.}\ \bibnamefont {Sarma}},\ }\href
	{https://doi.org/10.1103/PhysRevLett.125.117206} {\bibfield  {journal}
		{\bibinfo  {journal} {Phys. Rev. Lett.}\ }\textbf {\bibinfo {volume} {125}},\
		\bibinfo {pages} {117206} (\bibinfo {year} {2020})}\BibitemShut {NoStop}%
	\bibitem [{\citenamefont {Bernu}\ and\ \citenamefont
		{Misguich}(2001)}]{Bernu134409}%
	\BibitemOpen
	\bibfield  {author} {\bibinfo {author} {\bibfnamefont {B.}~\bibnamefont
			{Bernu}}\ and\ \bibinfo {author} {\bibfnamefont {G.}~\bibnamefont
			{Misguich}},\ }\href {https://doi.org/10.1103/PhysRevB.63.134409} {\bibfield
		{journal} {\bibinfo  {journal} {Phys. Rev. B}\ }\textbf {\bibinfo {volume}
			{63}},\ \bibinfo {pages} {134409} (\bibinfo {year} {2001})}\BibitemShut
	{NoStop}%
	\bibitem [{\citenamefont {Makivi\ifmmode~\acute{c}\else \'{c}\fi{}}\ and\
		\citenamefont {Ding}(1991)}]{Makivic3562}%
	\BibitemOpen
	\bibfield  {author} {\bibinfo {author} {\bibfnamefont {M.~S.}\ \bibnamefont
			{Makivi\ifmmode~\acute{c}\else \'{c}\fi{}}}\ and\ \bibinfo {author}
		{\bibfnamefont {H.-Q.}\ \bibnamefont {Ding}},\ }\href
	{https://doi.org/10.1103/PhysRevB.43.3562} {\bibfield  {journal} {\bibinfo
			{journal} {Phys. Rev. B}\ }\textbf {\bibinfo {volume} {43}},\ \bibinfo
		{pages} {3562} (\bibinfo {year} {1991})}\BibitemShut {NoStop}%
	\bibitem [{\citenamefont {Kim}\ and\ \citenamefont {Troyer}(1998)}]{Kim2705}%
	\BibitemOpen
	\bibfield  {author} {\bibinfo {author} {\bibfnamefont {J.-K.}\ \bibnamefont
			{Kim}}\ and\ \bibinfo {author} {\bibfnamefont {M.}~\bibnamefont {Troyer}},\
	}\href {https://doi.org/10.1103/PhysRevLett.80.2705} {\bibfield  {journal}
		{\bibinfo  {journal} {Phys. Rev. Lett.}\ }\textbf {\bibinfo {volume} {80}},\
		\bibinfo {pages} {2705} (\bibinfo {year} {1998})}\BibitemShut {NoStop}%
	\bibitem [{\citenamefont {Avella}\ and\ \citenamefont
		{Mancini}(2013)}]{Avella2013}%
	\BibitemOpen
	\bibfield  {author} {\bibinfo {author} {\bibfnamefont {A.}~\bibnamefont
			{Avella}}\ and\ \bibinfo {author} {\bibfnamefont {F.}~\bibnamefont
			{Mancini}},\ }\href@noop {} {\emph {\bibinfo {title} {Strongly Correlated
				Systems}}},\ Vol.\ \bibinfo {volume} {176}\ (\bibinfo  {publisher}
	{Springer},\ \bibinfo {address} {Berlin},\ \bibinfo {year}
	{2013})\BibitemShut {NoStop}%
	\bibitem [{\citenamefont {Lee}\ \emph {et~al.}(2020)\citenamefont {Lee},
		\citenamefont {Klauer}, \citenamefont {Menten}, \citenamefont {Lee},
		\citenamefont {Yoon}, \citenamefont {Luetkens}, \citenamefont {Lemmens},
		\citenamefont {M\"oller},\ and\ \citenamefont {Choi}}]{Lee224420}%
	\BibitemOpen
	\bibfield  {author} {\bibinfo {author} {\bibfnamefont {S.}~\bibnamefont
			{Lee}}, \bibinfo {author} {\bibfnamefont {R.}~\bibnamefont {Klauer}},
		\bibinfo {author} {\bibfnamefont {J.}~\bibnamefont {Menten}}, \bibinfo
		{author} {\bibfnamefont {W.}~\bibnamefont {Lee}}, \bibinfo {author}
		{\bibfnamefont {S.}~\bibnamefont {Yoon}}, \bibinfo {author} {\bibfnamefont
			{H.}~\bibnamefont {Luetkens}}, \bibinfo {author} {\bibfnamefont
			{P.}~\bibnamefont {Lemmens}}, \bibinfo {author} {\bibfnamefont
			{A.}~\bibnamefont {M\"oller}},\ and\ \bibinfo {author} {\bibfnamefont
			{K.-Y.}\ \bibnamefont {Choi}},\ }\href
	{https://doi.org/10.1103/PhysRevB.101.224420} {\bibfield  {journal} {\bibinfo
			{journal} {Phys. Rev. B}\ }\textbf {\bibinfo {volume} {101}},\ \bibinfo
		{pages} {224420} (\bibinfo {year} {2020})}\BibitemShut {NoStop}%
	\bibitem [{\citenamefont {Uemura}\ \emph {et~al.}(1994)\citenamefont {Uemura},
		\citenamefont {Keren}, \citenamefont {Kojima}, \citenamefont {Le},
		\citenamefont {Luke}, \citenamefont {Wu}, \citenamefont {Ajiro},
		\citenamefont {Asano}, \citenamefont {Kuriyama}, \citenamefont {Mekata},
		\citenamefont {Kikuchi},\ and\ \citenamefont {Kakurai}}]{Uemura3306}%
	\BibitemOpen
	\bibfield  {author} {\bibinfo {author} {\bibfnamefont {Y.~J.}\ \bibnamefont
			{Uemura}}, \bibinfo {author} {\bibfnamefont {A.}~\bibnamefont {Keren}},
		\bibinfo {author} {\bibfnamefont {K.}~\bibnamefont {Kojima}}, \bibinfo
		{author} {\bibfnamefont {L.~P.}\ \bibnamefont {Le}}, \bibinfo {author}
		{\bibfnamefont {G.~M.}\ \bibnamefont {Luke}}, \bibinfo {author}
		{\bibfnamefont {W.~D.}\ \bibnamefont {Wu}}, \bibinfo {author} {\bibfnamefont
			{Y.}~\bibnamefont {Ajiro}}, \bibinfo {author} {\bibfnamefont
			{T.}~\bibnamefont {Asano}}, \bibinfo {author} {\bibfnamefont
			{Y.}~\bibnamefont {Kuriyama}}, \bibinfo {author} {\bibfnamefont
			{M.}~\bibnamefont {Mekata}}, \bibinfo {author} {\bibfnamefont
			{H.}~\bibnamefont {Kikuchi}},\ and\ \bibinfo {author} {\bibfnamefont
			{K.}~\bibnamefont {Kakurai}},\ }\href
	{https://doi.org/10.1103/PhysRevLett.73.3306} {\bibfield  {journal} {\bibinfo
			{journal} {Phys. Rev. Lett.}\ }\textbf {\bibinfo {volume} {73}},\ \bibinfo
		{pages} {3306} (\bibinfo {year} {1994})}\BibitemShut {NoStop}%
	\bibitem [{\citenamefont {Lee}\ \emph {et~al.}(2021)\citenamefont {Lee},
		\citenamefont {Lee}, \citenamefont {Berlie}, \citenamefont {Hillier},
		\citenamefont {Adroja}, \citenamefont {Zhong}, \citenamefont {Cava},
		\citenamefont {Jang},\ and\ \citenamefont {Choi}}]{Lee024413}%
	\BibitemOpen
	\bibfield  {author} {\bibinfo {author} {\bibfnamefont {S.}~\bibnamefont
			{Lee}}, \bibinfo {author} {\bibfnamefont {C.~H.}\ \bibnamefont {Lee}},
		\bibinfo {author} {\bibfnamefont {A.}~\bibnamefont {Berlie}}, \bibinfo
		{author} {\bibfnamefont {A.~D.}\ \bibnamefont {Hillier}}, \bibinfo {author}
		{\bibfnamefont {D.~T.}\ \bibnamefont {Adroja}}, \bibinfo {author}
		{\bibfnamefont {R.}~\bibnamefont {Zhong}}, \bibinfo {author} {\bibfnamefont
			{R.~J.}\ \bibnamefont {Cava}}, \bibinfo {author} {\bibfnamefont {Z.~H.}\
			\bibnamefont {Jang}},\ and\ \bibinfo {author} {\bibfnamefont {K.-Y.}\
			\bibnamefont {Choi}},\ }\href {https://doi.org/10.1103/PhysRevB.103.024413}
	{\bibfield  {journal} {\bibinfo  {journal} {Phys. Rev. B}\ }\textbf {\bibinfo
			{volume} {103}},\ \bibinfo {pages} {024413} (\bibinfo {year}
		{2021})}\BibitemShut {NoStop}%
	\bibitem [{\citenamefont {Keren}\ \emph {et~al.}(1996)\citenamefont {Keren},
		\citenamefont {Mendels}, \citenamefont {Campbell},\ and\ \citenamefont
		{Lord}}]{Keren1386}%
	\BibitemOpen
	\bibfield  {author} {\bibinfo {author} {\bibfnamefont {A.}~\bibnamefont
			{Keren}}, \bibinfo {author} {\bibfnamefont {P.}~\bibnamefont {Mendels}},
		\bibinfo {author} {\bibfnamefont {I.~A.}\ \bibnamefont {Campbell}},\ and\
		\bibinfo {author} {\bibfnamefont {J.}~\bibnamefont {Lord}},\ }\href
	{https://doi.org/10.1103/PhysRevLett.77.1386} {\bibfield  {journal} {\bibinfo
			{journal} {Phys. Rev. Lett.}\ }\textbf {\bibinfo {volume} {77}},\ \bibinfo
		{pages} {1386} (\bibinfo {year} {1996})}\BibitemShut {NoStop}%
	\bibitem [{\citenamefont {Onishi}\ \emph {et~al.}(2012)\citenamefont {Onishi},
		\citenamefont {Oka}, \citenamefont {Azuma}, \citenamefont {Shimakawa},
		\citenamefont {Motome}, \citenamefont {Taniguchi}, \citenamefont {Hiraishi},
		\citenamefont {Miyazaki}, \citenamefont {Masuda}, \citenamefont {Koda},
		\citenamefont {Kojima},\ and\ \citenamefont {Kadono}}]{Onishi184412}%
	\BibitemOpen
	\bibfield  {author} {\bibinfo {author} {\bibfnamefont {N.}~\bibnamefont
			{Onishi}}, \bibinfo {author} {\bibfnamefont {K.}~\bibnamefont {Oka}},
		\bibinfo {author} {\bibfnamefont {M.}~\bibnamefont {Azuma}}, \bibinfo
		{author} {\bibfnamefont {Y.}~\bibnamefont {Shimakawa}}, \bibinfo {author}
		{\bibfnamefont {Y.}~\bibnamefont {Motome}}, \bibinfo {author} {\bibfnamefont
			{T.}~\bibnamefont {Taniguchi}}, \bibinfo {author} {\bibfnamefont
			{M.}~\bibnamefont {Hiraishi}}, \bibinfo {author} {\bibfnamefont
			{M.}~\bibnamefont {Miyazaki}}, \bibinfo {author} {\bibfnamefont
			{T.}~\bibnamefont {Masuda}}, \bibinfo {author} {\bibfnamefont
			{A.}~\bibnamefont {Koda}}, \bibinfo {author} {\bibfnamefont {K.~M.}\
			\bibnamefont {Kojima}},\ and\ \bibinfo {author} {\bibfnamefont
			{R.}~\bibnamefont {Kadono}},\ }\href
	{https://doi.org/10.1103/PhysRevB.85.184412} {\bibfield  {journal} {\bibinfo
			{journal} {Phys. Rev. B}\ }\textbf {\bibinfo {volume} {85}},\ \bibinfo
		{pages} {184412} (\bibinfo {year} {2012})}\BibitemShut {NoStop}%
	\bibitem [{\citenamefont {Campbell}\ \emph {et~al.}(1994)\citenamefont
		{Campbell}, \citenamefont {Amato}, \citenamefont {Gygax}, \citenamefont
		{Herlach}, \citenamefont {Schenck}, \citenamefont {Cywinski},\ and\
		\citenamefont {Kilcoyne}}]{Campbell1291}%
	\BibitemOpen
	\bibfield  {author} {\bibinfo {author} {\bibfnamefont {I.~A.}\ \bibnamefont
			{Campbell}}, \bibinfo {author} {\bibfnamefont {A.}~\bibnamefont {Amato}},
		\bibinfo {author} {\bibfnamefont {F.~N.}\ \bibnamefont {Gygax}}, \bibinfo
		{author} {\bibfnamefont {D.}~\bibnamefont {Herlach}}, \bibinfo {author}
		{\bibfnamefont {A.}~\bibnamefont {Schenck}}, \bibinfo {author} {\bibfnamefont
			{R.}~\bibnamefont {Cywinski}},\ and\ \bibinfo {author} {\bibfnamefont
			{S.~H.}\ \bibnamefont {Kilcoyne}},\ }\href
	{https://doi.org/10.1103/PhysRevLett.72.1291} {\bibfield  {journal} {\bibinfo
			{journal} {Phys. Rev. Lett.}\ }\textbf {\bibinfo {volume} {72}},\ \bibinfo
		{pages} {1291} (\bibinfo {year} {1994})}\BibitemShut {NoStop}%
	\bibitem [{\citenamefont {Mendels}\ \emph {et~al.}(2007)\citenamefont
		{Mendels}, \citenamefont {Bert}, \citenamefont {de~Vries}, \citenamefont
		{Olariu}, \citenamefont {Harrison}, \citenamefont {Duc}, \citenamefont
		{Trombe}, \citenamefont {Lord}, \citenamefont {Amato},\ and\ \citenamefont
		{Baines}}]{Mendels077204}%
	\BibitemOpen
	\bibfield  {author} {\bibinfo {author} {\bibfnamefont {P.}~\bibnamefont
			{Mendels}}, \bibinfo {author} {\bibfnamefont {F.}~\bibnamefont {Bert}},
		\bibinfo {author} {\bibfnamefont {M.~A.}\ \bibnamefont {de~Vries}}, \bibinfo
		{author} {\bibfnamefont {A.}~\bibnamefont {Olariu}}, \bibinfo {author}
		{\bibfnamefont {A.}~\bibnamefont {Harrison}}, \bibinfo {author}
		{\bibfnamefont {F.}~\bibnamefont {Duc}}, \bibinfo {author} {\bibfnamefont
			{J.~C.}\ \bibnamefont {Trombe}}, \bibinfo {author} {\bibfnamefont {J.~S.}\
			\bibnamefont {Lord}}, \bibinfo {author} {\bibfnamefont {A.}~\bibnamefont
			{Amato}},\ and\ \bibinfo {author} {\bibfnamefont {C.}~\bibnamefont
			{Baines}},\ }\href {https://doi.org/10.1103/PhysRevLett.98.077204} {\bibfield
		{journal} {\bibinfo  {journal} {Phys. Rev. Lett.}\ }\textbf {\bibinfo
			{volume} {98}},\ \bibinfo {pages} {077204} (\bibinfo {year}
		{2007})}\BibitemShut {NoStop}%
	\bibitem [{\citenamefont {Guo}\ \emph {et~al.}(2020{\natexlab{b}})\citenamefont
		{Guo}, \citenamefont {Zhong}, \citenamefont {Górnicka}, \citenamefont
		{Klimczuk},\ and\ \citenamefont {Cava}}]{Guo10670}%
	\BibitemOpen
	\bibfield  {author} {\bibinfo {author} {\bibfnamefont {S.}~\bibnamefont
			{Guo}}, \bibinfo {author} {\bibfnamefont {R.}~\bibnamefont {Zhong}}, \bibinfo
		{author} {\bibfnamefont {K.}~\bibnamefont {Górnicka}}, \bibinfo {author}
		{\bibfnamefont {T.}~\bibnamefont {Klimczuk}},\ and\ \bibinfo {author}
		{\bibfnamefont {R.~J.}\ \bibnamefont {Cava}},\ }\href
	{https://doi.org/10.1021/acs.chemmater.0c03850} {\bibfield  {journal}
		{\bibinfo  {journal} {Chem. Mater}\ }\textbf {\bibinfo {volume} {32}},\
		\bibinfo {pages} {10670} (\bibinfo {year} {2020}{\natexlab{b}})}\BibitemShut
	{NoStop}%
	\bibitem [{\citenamefont {Pan}\ \emph {et~al.}(2021)\citenamefont {Pan},
		\citenamefont {Ni}, \citenamefont {He}, \citenamefont {Yu}, \citenamefont
		{Xu},\ and\ \citenamefont {Li}}]{Pan104412}%
	\BibitemOpen
	\bibfield  {author} {\bibinfo {author} {\bibfnamefont {B.~L.}\ \bibnamefont
			{Pan}}, \bibinfo {author} {\bibfnamefont {J.~M.}\ \bibnamefont {Ni}},
		\bibinfo {author} {\bibfnamefont {L.~P.}\ \bibnamefont {He}}, \bibinfo
		{author} {\bibfnamefont {Y.~J.}\ \bibnamefont {Yu}}, \bibinfo {author}
		{\bibfnamefont {Y.}~\bibnamefont {Xu}},\ and\ \bibinfo {author}
		{\bibfnamefont {S.~Y.}\ \bibnamefont {Li}},\ }\href
	{https://doi.org/10.1103/PhysRevB.103.104412} {\bibfield  {journal} {\bibinfo
			{journal} {Phys. Rev. B}\ }\textbf {\bibinfo {volume} {103}},\ \bibinfo
		{pages} {104412} (\bibinfo {year} {2021})}\BibitemShut {NoStop}%
	\bibitem [{\citenamefont {Rauch}\ \emph {et~al.}(2015)\citenamefont {Rauch},
		\citenamefont {Kraken}, \citenamefont {Litterst}, \citenamefont {S\"ullow},
		\citenamefont {Luetkens}, \citenamefont {Brando}, \citenamefont {F\"orster},
		\citenamefont {Sichelschmidt}, \citenamefont {Neubauer}, \citenamefont
		{Pfleiderer}, \citenamefont {Duncan},\ and\ \citenamefont
		{Grosche}}]{Rauch174404}%
	\BibitemOpen
	\bibfield  {author} {\bibinfo {author} {\bibfnamefont {D.}~\bibnamefont
			{Rauch}}, \bibinfo {author} {\bibfnamefont {M.}~\bibnamefont {Kraken}},
		\bibinfo {author} {\bibfnamefont {F.~J.}\ \bibnamefont {Litterst}}, \bibinfo
		{author} {\bibfnamefont {S.}~\bibnamefont {S\"ullow}}, \bibinfo {author}
		{\bibfnamefont {H.}~\bibnamefont {Luetkens}}, \bibinfo {author}
		{\bibfnamefont {M.}~\bibnamefont {Brando}}, \bibinfo {author} {\bibfnamefont
			{T.}~\bibnamefont {F\"orster}}, \bibinfo {author} {\bibfnamefont
			{J.}~\bibnamefont {Sichelschmidt}}, \bibinfo {author} {\bibfnamefont
			{A.}~\bibnamefont {Neubauer}}, \bibinfo {author} {\bibfnamefont
			{C.}~\bibnamefont {Pfleiderer}}, \bibinfo {author} {\bibfnamefont {W.~J.}\
			\bibnamefont {Duncan}},\ and\ \bibinfo {author} {\bibfnamefont {F.~M.}\
			\bibnamefont {Grosche}},\ }\href {https://doi.org/10.1103/PhysRevB.91.174404}
	{\bibfield  {journal} {\bibinfo  {journal} {Phys. Rev. B}\ }\textbf {\bibinfo
			{volume} {91}},\ \bibinfo {pages} {174404} (\bibinfo {year}
		{2015})}\BibitemShut {NoStop}%
	\bibitem [{\citenamefont {Abragam}\ and\ \citenamefont
		{Bleaney}(2012)}]{Abragam2012}%
	\BibitemOpen
	\bibfield  {author} {\bibinfo {author} {\bibfnamefont {A.}~\bibnamefont
			{Abragam}}\ and\ \bibinfo {author} {\bibfnamefont {B.}~\bibnamefont
			{Bleaney}},\ }\href@noop {} {\emph {\bibinfo {title} {{Electron paramagnetic
					resonance of transition ions}}}}\ (\bibinfo  {publisher} {OUP Oxford},\
	\bibinfo {year} {2012})\BibitemShut {NoStop}%
	\bibitem [{\citenamefont {Amato}\ \emph {et~al.}(2017)\citenamefont {Amato},
		\citenamefont {Luetkens}, \citenamefont {Sedlak}, \citenamefont {Stoykov},
		\citenamefont {Scheuermann}, \citenamefont {Elender}, \citenamefont
		{Raselli},\ and\ \citenamefont {Graf}}]{Amato093301}%
	\BibitemOpen
	\bibfield  {author} {\bibinfo {author} {\bibfnamefont {A.}~\bibnamefont
			{Amato}}, \bibinfo {author} {\bibfnamefont {H.}~\bibnamefont {Luetkens}},
		\bibinfo {author} {\bibfnamefont {K.}~\bibnamefont {Sedlak}}, \bibinfo
		{author} {\bibfnamefont {A.}~\bibnamefont {Stoykov}}, \bibinfo {author}
		{\bibfnamefont {R.}~\bibnamefont {Scheuermann}}, \bibinfo {author}
		{\bibfnamefont {M.}~\bibnamefont {Elender}}, \bibinfo {author} {\bibfnamefont
			{A.}~\bibnamefont {Raselli}},\ and\ \bibinfo {author} {\bibfnamefont
			{D.}~\bibnamefont {Graf}},\ }\href {https://doi.org/10.1063/1.4986045}
	{\bibfield  {journal} {\bibinfo  {journal} {Rev. Sci. Instrum.}\ }\textbf
		{\bibinfo {volume} {88}},\ \bibinfo {pages} {093301} (\bibinfo {year}
		{2017})}\BibitemShut {NoStop}%
	\bibitem [{\citenamefont {Orbach}\ and\ \citenamefont
		{Bleaney}(1961)}]{Orbach458}%
	\BibitemOpen
	\bibfield  {author} {\bibinfo {author} {\bibfnamefont {R.~.}\ \bibnamefont
			{Orbach}}\ and\ \bibinfo {author} {\bibfnamefont {B.}~\bibnamefont
			{Bleaney}},\ }\href {https://doi.org/10.1098/rspa.1961.0211} {\bibfield
		{journal} {\bibinfo  {journal} {Proc. Math. Phys. Eng. Sci.}\ }\textbf
		{\bibinfo {volume} {264}},\ \bibinfo {pages} {458} (\bibinfo {year}
		{1961})}\BibitemShut {NoStop}%
\end{thebibliography}
%

\end{document}